\documentclass[showpacs,preprintnumbers,amsmath,amssymb]{revtex4}

\usepackage{graphicx}% Include figure files
\usepackage{dcolumn}% Align table columns on decimal point
\usepackage{bm}% bold math
\begin{document}

\title{Quantum master equation for a system influencing its environment}

\author{Massimiliano Esposito}
\author{Pierre Gaspard}
\affiliation{Center for Nonlinear Phenomena and Complex Systems,\\
Universit\'e Libre de Bruxelles, Code Postal 231, Campus Plaine, B-1050
Brussels, Belgium.}

\date{\today}
\pacs{05.50.+q, 03.65.Yz}
%%%%%%%%%%%%%%%%%%%%%%%%%%%%%%%%%%%%%%%%%%%%%%%%%%%%%%%%%%%%%%%%%%%%%%%%%%%%%%%%

\begin{abstract}
A perturbative quantum master equation is derived for a system
interacting with its environment, which is more general than
the ones derived before.  Our master equation takes into account
the effect of the energy exchanges between the
system and the environment and the conservation of energy in
the finite total system.  This master equation describes
relaxation mechanisms in isolated nanoscopic quantum systems.
In its most general form, this equation is non-Markovian and a
Markovian version of it rules the long-time relaxation.
We show that our equation reduces to the Redfield equation
in the limit where the energy of the system does not
affect the density of state of its environment. This master equation
and the Redfield one are applied to a spin-environment model
defined in terms of random matrices and
compared with the solutions of the exact von Neumann equation.
The comparison proves the necessity to allow
energy exchange between the subsystem and the environment
in order to correctly describe the relaxation
in an isolated nanoscopic total system.
\end{abstract}

\maketitle
%%%%%%%%%%%%%%%%%%%%%%%%%%%%%%%%%%%%%%%%%%%%%%%%%%%%%%%%%%%%%%%%%%%%%%%%%%%%
\section{Introduction \label{intro}}

Studying the dynamics of a simple system interacting with its
environment is a very important problem in physics. The
theoretical description of this problem started a long time ago.

In the context of classical mechanics several master equations,
such as the Boltzmann equation, the Chapman-Kolmogorov master equation
or the Fokker-Planck equation, were derived in order to describe
the time evolution of the probability density of the system variables.

In the context of quantum mechanics,
wich interests us in this paper, the time evolution of a system interacting
with its environment is described
in terms of a reduced density matrix that is obtained by tracing out
the degrees of freedom of the environment from the total (system
plus environment) density matrix. In this way, the first quantum master
equation was obtained by Pauli \cite{Pauli,Zwanzig,Zubarev} in $1928$. This
equation is called the \textit{Pauli equation} and describes the
evolution of the populations (i.e., the diagonal elements of the density
matrix) when the system is weakly perturbed by an additional term
in its Hamiltonian. The transition rates between populations are
given by the Fermi golden rule. In $1957$, Redfield \cite{Red}
derived the so-called \textit{Redfield equation} in
the context of NMR for a system such as a spin interacting with its
environment.
This equation has been widely used and applied to many systems where
the dynamics of the environment is faster than the dynamics of the system.
This equation is Markovian and has the defect of breaking the
positivity on short time scales of the order of the environment 
correlation time for initial conditions near the border of the space of
physically admissible density matrices. Many similar master equations 
for a system interacting with an environment have been derived since then
starting from
the von Neumann equation and making several assumptions (weak
coupling limit, Markovianity, separation of time scale between
system and environment) \cite{Tannoudji,Gardiner,Kampen,Haake1,Weiss}.
In $1976$ Lindblad \cite{Lindblad} derived the most general quantum
master equation which is Markovian and which preserves positivity.
The Redfield equation has a Lindblad form in the case of
$\delta$-correlated environments.
More recently, a \textit{non-Markovian Redfield equation} has been obtained
that preserves positivity and reduces to the Redfield equation in the
Markovian limit \cite{GaspRed}.
It has also been shown \cite{GaspRed,Suarez} that the Markovian Redfield equation
can preserve positivity if one applies a slippage of initial conditions that
takes into account the non-Markovian effects on the early
dynamics. Similar considerations have been proposed for different master equations
\cite{Yu-Struntz,Struntz,Haake2}.
As far as one considers the weak-coupling regime, all the master equations 
derived till now in the literature at second order of perturbation theory 
can be deduced from the non-Markovian Redfield equation.

The problem is that the non-Markovian Redfield equation
as well as the other aforementioned master equations existing in the literature
are based on the fundamental assumption that the environment does not feel the
effect of the system. This assumption seems realistic for
macroscopic environments but not in the case of nanoscopic isolated total systems
in which the density of states of the environment can vary on an energy scale of 
the order of the system energy scale. Because nanoscopic physics is 
experimentally progressing very fast, we expect that such effects will become
important and measurable in future applications. Already, quantum dissipation
is being envisaged on the nanoscale for applications
such as spin dynamics in quantum dots \cite{Cohen} or isomerizations in
atomic or molecular clusters
in microcanonical statistical ensembles \cite{Marcus}.
Another possible application is the intramolecular energy relaxation
in polyatomic molecules \cite{Rice}.

The aim of the present paper is to systematically derive from the von Neumann
equation a master equation which takes into account the fact that
the energy of the total system (system plus environment) is finite
and constant and, therefore, that the energy distribution of the environment
is affected by energy exchanges with the
system. The aforementioned equations can be derived from our master
equation,
which thus appears to be very general.

The plan of the paper is the following. In Sec. \ref{genpauleq} we
systematically derive our master equation and the non-Markovian Redfield
equation
from the von Neumann equation, by perfoming a second-order perturbative
expansion
in the coupling parameter (under the assumption of weak coupling) for
general environments.
Thereafter, we consider the Markovian limit in
both cases. We also show how, in this limit and
neglecting the coupling between the populations and
the quantum coherences, our master equation reduces to a simple equation of
Pauli type for the total system, taking into account the
modifications of the energy distribution of the environment due to
the energy exchanges with the system. Finally, we compare
our master equation to the Redfield equation and discuss how the
Redfield equation can be seen as a particular case
of our master equation. In Sec. \ref{spin-boson} we apply 
our master equation and the Redfield equation to the case
where the system is a two-level system interacting with a general
environment. In Sec. \ref{spin-GORM}, we apply the master equations
to the case where the system is a two-level system interacting
with a complex environment (such as a classically chaotic or many-body
environment)
that is modeled by random matrices from a Gaussian orthogonal ensemble,
which we call
Gaussian orthogonal random matrices (GORM).
In Sec. \ref{numres}, we compare the solutions
of the non-Markovian and Markovian master equations to
the exact solutions of the complete von
Neumann equation in the case of our spin-GORM model.
Conclusions are finally drawn in Sec. \ref{conc}.

%%%%%%%%%%%%%%%%%%%%%%%%%%%%%%%%%%%%%%%%%%%%%%%%%%%%%%%%%%%%%%%%%%%%%%%%%%%%%%%%
%
%%%%%%%%%%%%%%%%%%%%%%%%%%%%%%%%%%%%%%%%%%%%%%%%%%%%%%%%%%%%%%%%%%%%%%%%%%%%%%%%
%
\section{Derivation of the fundamental equations \label{genpauleq}}

The Hamiltonian of the total systems that we consider here is made
of the sum of the system Hamiltonian $\hat{H}_{S}$ and the
environment Hamiltonian $\hat{H}_{B}$ plus a coupling term that
has the form of the product of a system operator $\hat{S}$ and a
environment operator $\hat{B}$. The generalization to a coupling
term of the form $\lambda \sum_i \hat{S_i} \hat{B_i}$ is easy. The
amplitude of the coupling term is determined by the coupling
parameter $\lambda$:
\begin{eqnarray}
\hat{H}_{\rm tot}&=&\hat{H}_{0}+\lambda \hat{V} \nonumber \\
&=&\hat{H}_{S}+\hat{H}_{B}+\lambda \hat{S} \hat{B} \; .
\label{hamiltonientotal}
\end{eqnarray}
The eigenstates of $\hat{H}_{S}$, respectively of $\hat{H}_B$,
will be denoted by $\vert s \rangle$, respectively $\vert b
\rangle$. The eigenvalues of $\hat{H}_{S}$, respectively of
$\hat{H}_B$, will be denoted by $E_s$, respectively $E_b$.
Finally, the eigenstates of $\hat{H}_{\rm tot}$ will be denoted by
$\vert \alpha \rangle$ and its eigenvalues by $E_{\alpha}$.\\
The evolution of the total density matrix is described by the von
Neumann equation:
\begin{equation}
\dot{\hat{\rho}}(t)=-i\lbrack\hat{H}_{\rm tot},\hat{\rho}(t)\rbrack \equiv
{\cal L}_{\rm tot}\hat{\rho}(t) \; ,
\label{vonNeumann}
\end{equation}
where ${\cal L}_{\rm tot}$ is the so-called quantum Liouvillian or von
Neumann operator of the total system.
The interaction representations of the operators are given by
\begin{eqnarray}
\hat{\rho}_{I}(t)&=&e^{i\hat{H}_{0}t}\hat{\rho}(t)e^{-i\hat{H}_{0}t} \;
,\nonumber \\
\hat{V}(t)&=&e^{i\hat{H}_{0}t}\hat{V}e^{-i\hat{H}_{0}t} \; ,\nonumber \\
\hat{B}(t)&=&e^{i\hat{H}_{B}t}\hat{B}e^{-i\hat{H}_{B}t} \; ,\nonumber \\
\hat{S}(t)&=&e^{i\hat{H}_{S}t}\hat{S}e^{-i\hat{H}_{S}t} \; .
\label{formeinteraction}
\end{eqnarray}
In the interaction representation, the von Neumann equation becomes
\begin{equation}
\dot{\hat{\rho}}_I(t)=-i \lbrack \lambda
\hat{V}(t),\hat{\rho}_I(t) \rbrack \equiv{\cal L}_I(t)\hat{\rho}_I(t) \; ,
\label{vonNeumanninteraction}
\end{equation}
with the interaction Liouvillian ${\cal L}_I(t)=e^{-{\cal L}_0t}{\cal L}_I
e^{{\cal L}_0t}$ where $e^{{\cal L}_0 t} \hat{A} = e^{-i\hat{H}_{0}t} 
\hat{A} e^{i\hat{H}_{0}t}$, the free Liouvillian ${\cal L}_0={\cal L}_S+
{\cal L}_B=-i\lbrack \hat{H}_S,\cdot \rbrack-i\lbrack \hat{H}_B,\cdot 
\rbrack$, and $\hat{A}$ is an arbitrary operator.
The perturbative expression of the von Neumann equation in the
interaction representation is given to order $\lambda^2$ by
\begin{eqnarray}
\hat{\rho}_{I}(t)
%&=&\hat{\rho}(0)
%-i \lambda \int^{t}_{0} dt_1 \lbrack \hat{V}(t_1),\hat{\rho}(0)\rbrack
%\nonumber \\
%& &- \lambda^2 \int^{t}_{0} dt_1 \int^{t_1}_{0} dt_2 \lbrack
%\hat{V}(t_1),\lbrack \hat{V}(t_2)
%,\hat{\rho}(0)\rbrack\rbrack + O(\lambda^3) \\
&=&\hat{\rho}(0)
+ \int^{t}_{0} dt_1 {\cal L}_I(t_1)\hat{\rho}(0) \nonumber \\
& &+ \int^{t}_{0} dt_1 \int^{t_1}_{0} dt_2 {\cal L}_I(t_1){\cal
L}_I(t_2)\hat\rho(0)
+ O(\lambda^3) \\
&=&\hat{\rho}(0) + \int^{t}_{0} dT e^{-{\cal L}_0T}{\cal L}_Ie^{{\cal
L}_0T}\hat{\rho}(0) \nonumber \\
& &+ \int^{t}_{0} dT \int^{T}_{0} d\tau e^{-{\cal L}_0T}{\cal L}_Ie^{{\cal
L}_0\tau}{\cal
L}_Ie^{-{\cal L}_0\tau}e^{{\cal L}_0T}\hat\rho(0)  + O(\lambda^3) \; ,
\label{rhototintpertrurb}
\end{eqnarray}
if we set $T=t_1$ and $\tau=t_1-t_2$.
Equation (\ref{rhototintpertrurb}) is the starting point of all the
derivations of a
master equation in the weak coupling limit for a total system made
of a system and its environment in mutual interaction.

%%%%%%%%%%%%%%%%%%%%%%%%%%%%%%%%%%%%%%%%%%%%%%%%%%%%%%%%%%%%%%%%%%%%%%%%%%%%%%%%
%
\subsection{Our quantum master equation}

We now derive our master equation which is the central result of
this paper.

The main idea is to describe the time evolution in terms of quantities
which are distributed over the energy of the environment.
We thus define the following quantities in terms of which we intend to describe
the properties of the system:
\begin{equation}
P_{ss'}(\epsilon;t) \equiv {\rm Tr} \; \hat\rho(t) \vert s'\rangle\langle s
\vert \; \delta(\epsilon-\hat
H_B)\; .
\label{Pss'}
\end{equation}
The diagonal element $P_{ss}(\epsilon;t)$ is the probability density to
find the system in the
state $s$ while the environment has the energy $\epsilon$.  The
off-diagonal element
$P_{ss'}(\epsilon;t)$ characterizes the density of the quantum coherence
between the states $s$ and $s'$,
density which is distributed over the energy $\epsilon$ of the environment.

The matrix composed of the elements $P_{ss'}(\epsilon;t)$ is Hermitian
\begin{equation}
P_{ss'}(\epsilon;t)=P_{s's}^{\ast}(\epsilon;t) \; .
\label{Phermi}
\end{equation}

Moreover, the normalization ${\rm Tr}\hat\rho(t)=1$ of the total density
matrix implies that
\begin{equation}
\sum_s \int d\epsilon \; P_{ss}(\epsilon;t)=1 \; .
\label{norm}
\end{equation}

In order to obtain a closed description in terms of the quantities
(\ref{Pss'}),
we suppose that the total density matrix can be described at all times by a
density matrix of the
following form:
\begin{equation}
\hat{\rho}(t)= \sum_{s,s'} \vert s \rangle \langle s' \vert \;
\frac{P_{ss'}(\hat{H}_B;t)}{n(\hat H_B)},
\label{formegenrhopauli}
\end{equation}
where we have defined the energy density
\begin{equation}
n(\epsilon)=\textrm{Tr}_{B} \delta(\epsilon-\hat{H}_B),
\end{equation}
which is supposed to be smoothened on the energy scale
of the mean level spacing.
The assumption (\ref{formegenrhopauli}) has the effect of neglecting the
contributions from
the environment coherences to the system dynamics (albeit the system
coherences are kept in the description).
We remark that the form (\ref{formegenrhopauli}) is not supposed to
strictly hold at all times but is an assumption
in order to obtain a closed set of equations for the quantities
$P_{ss'}(\epsilon;t)$.

In order to better understand the meaning of the above definitions, we
notice that the reduced density matrix
of the system takes the form
\begin{equation}
\hat{\rho}_S(t)=\textrm{Tr}_{B} \hat{\rho}(t)=
\int d\epsilon \textrm{Tr}_{B} \delta(\epsilon-\hat{H}_B) \hat{\rho}(t)=
\int d\epsilon \sum_{s,s'} \vert s \rangle P_{ss'}(\epsilon;t) \langle s'
\vert  \; ,\label{rhospaulimic}
\end{equation}
which can be represented in the basis of the eigenstates of the system
Hamiltonian as
\begin{equation}
\hat{\rho}_S(t)=\int d\epsilon
\left(\begin{array}{cccc}
P_{11}(\epsilon;t) & P_{12}(\epsilon;t) & \ldots & P_{1N_S}(\epsilon;t) \\
P_{21}(\epsilon;t) & P_{22}(\epsilon;t) & \ldots & P_{2N_S}(\epsilon;t) \\
\vdots & \vdots & \ddots & \vdots \\
P_{N_S1}(\epsilon;t) & P_{N_S 2}(\epsilon;t) & \ldots &
P_{N_SN_S}(\epsilon;t) \\
\end{array} \right) \; . \label{rhoSexplicite}
\end{equation}
The goal of the precedent choice for the form of $\hat{\rho}(t)$ is thus
to obtain a description in which the state $s$ of the system is
correlated with the energy $\epsilon$ of the environment. In other words,
the density
matrix $\hat\rho_S$ of the system is decomposed as a distribution over the
energy $\epsilon$ of the
environment.

We now proceed to the derivation of the equations of motion for our
quantities $P_{s s'}(\epsilon;t)$
in the weak-coupling limit.  We start from the perturbative expansion
(\ref{rhototintpertrurb})
of the total density matrix in the interaction representation
(\ref{formeinteraction}).
We first define the interaction representation of our quantities (\ref{Pss'}):
\begin{equation}
P_{Iss'}(\epsilon;t)=e^{i(E_s-E_{s'})t} \; P_{ss'}(\epsilon;t) \; . \label{Pint1}
\end{equation}
We now have that
\begin{equation}
P_{Iss'}(\epsilon;t)= {\rm Tr} \; \hat\rho_I(t) \vert s'\rangle\langle s
\vert \; \delta(\epsilon-\hat H_B)\; .\label{Pint2}
\end{equation}
Inserting the perturbative expansion
(\ref{rhototintpertrurb}), we get
\begin{eqnarray}
P_{Iss'}(\epsilon;t)&=&{\rm Tr} \hat X \hat{\rho}(0)
+ \int^{t}_{0} dT \; {\rm Tr} \hat X e^{-{\cal L}_0T}{\cal L}_Ie^{{\cal
L}_0T}\hat{\rho}(0) \nonumber \\ & &+ \int^{t}_{0} dT \int^{T}_{0} d\tau \;
{\rm Tr} \hat X e^{-{\cal
L}_0T}{\cal L}_Ie^{{\cal L}_0\tau}{\cal L}_Ie^{-{\cal L}_0\tau}e^{{\cal
L}_0T}\hat\rho(0)  +
O(\lambda^3) \; ,\nonumber \\
\label{Xintpertrurb}
\end{eqnarray}
where $\hat X=\vert s'\rangle\langle s \vert \; \delta(\epsilon-\hat H_B)$.
Differentiating with respect to time, we obtain the equation
\begin{eqnarray}
\dot P_{Iss'}(\epsilon;t)&=&  {\rm Tr} \hat X e^{-{\cal L}_0 t}{\cal
L}_Ie^{{\cal
L}_0t}\hat{\rho}(0) \nonumber \\ & &+ \int^{t}_{0} d\tau {\rm Tr} \hat X
e^{-{\cal
L}_0t}{\cal L}_Ie^{{\cal L}_0\tau}{\cal L}_Ie^{-{\cal L}_0\tau}e^{{\cal
L}_0t}\hat\rho(0)  +
O(\lambda^3) \; ,
\label{dotPIss'}
\end{eqnarray}
where the initial density matrix takes the assumed form
(\ref{formegenrhopauli}) with $t=0$.

The first term is thus explicitly given by
\begin{eqnarray}
 {\rm Tr} \hat X e^{-{\cal L}_0 t}{\cal L}_Ie^{{\cal L}_0t}\hat{\rho}(0)
&=& -i \lambda \; {\rm Tr}  \vert s'\rangle\langle s \vert \;
\delta(\epsilon-\hat H_B)
e^{i\hat H_0 t}\lbrack \hat V , e^{-i\hat H_0 t}\hat{\rho}(0)e^{i\hat H_0
t}\rbrack e^{-i\hat H_0 t}
\nonumber \\ &=& -i \lambda \sum_{\bar{s}} e^{i(E_s-E_{\bar{s}})t} \langle
s\vert \hat S \vert \bar{s}\rangle
P_{\bar{s}s'}(\epsilon;0) n(\epsilon) \langle \hat B\rangle_{\epsilon} 
\nonumber \\ & & +i \lambda \sum_{\bar{s}} e^{-i(E_{s'}-E_{\bar{s}})t} 
\langle \bar{s}\vert \hat S \vert s'\rangle
P_{s\bar{s}}(\epsilon;0) n(\epsilon) \langle \hat B\rangle_{\epsilon} \; ,
\end{eqnarray}
with the environment coupling operator $\hat B$ averaged over the
microcanonical state of the environment
\begin{equation}
 \langle \hat B\rangle_{\epsilon} \equiv \frac{{\rm Tr}
\delta(\epsilon-\hat H_B) \hat B}{n(\epsilon)} \; .
\end{equation}

We now assume that this average vanishes, $\langle \hat
B\rangle_{\epsilon}=0$.  Otherwise, the
nonvanishing average is absorbed in the system Hamiltonian by the
following substitutions:
\begin{eqnarray}
\hat H_S &\to& \hat H_S + \lambda \langle \hat B\rangle_{\epsilon} \hat S
\; , \\
\hat V &\to& \hat V - \langle \hat B\rangle_{\epsilon} \hat S \; , \\
\hat H_B &\to& \hat H_B \; ,
\end{eqnarray}
leaving unchanged the total Hamiltonian.  Thanks to this simplification, the
first-order term of the perturbative expansion vanishes
\begin{equation}
 {\rm Tr} \hat X e^{-{\cal L}_0 t}{\cal L}_Ie^{{\cal L}_0t}\hat{\rho}(0) =0
\; .
\end{equation}
As a consequence, the time evolution of the total density matrix is given
by the uncoupled
Hamiltonian $\hat H_0=\hat H_S+\hat H_B$ up to correction of the order of
$\lambda^2$:
\begin{equation}
\hat \rho(t) = e^{{\cal L}_0t } \hat \rho (0) + O(\lambda^2) \; .
\end{equation}
According to our closure assumption that the total density matrix keeps the
form (\ref{formegenrhopauli})
during its time evolution, we have that
\begin{equation}
e^{{\cal L}_0t } \hat \rho (0) = \hat{\rho}(t) + O(\lambda^2) =
\sum_{\bar{s},\bar{s}'} \vert \bar{s} \rangle
\langle \bar{s}'\vert \; \frac{P_{\bar{s}\bar{s}'}(\hat{H}_B;t)}{n(\hat
H_B)} + O(\lambda^2) \; ,
\end{equation}
which we can substitute in Eq. (\ref{dotPIss'}) to get
\begin{eqnarray}
\dot P_{Iss'}(\epsilon;t)&=&\int^{t}_{0} d\tau {\rm Tr} \hat X e^{-{\cal
L}_0t}{\cal L}_Ie^{{\cal L}_0\tau}{\cal L}_Ie^{-{\cal L}_0\tau}e^{{\cal
L}_0t}\hat\rho(0)  +
O(\lambda^3) \nonumber \\
&=& -\lambda^2 e^{i(E_s-E_{s'})t} \sum_{\bar{s},\bar{s}'} \int^{t}_{0}
d\tau \; {\rm Tr}\left\{
\vert s'\rangle\langle s \vert \; \delta(\epsilon-\hat H_B) \right.\nonumber\\
& & \left.\times \left\lbrack \hat S \hat B ,
\left\lbrack \hat S(-\tau) \hat B(-\tau) , \vert \bar{s} \rangle\langle
\bar{s}' \vert
\frac{P_{\bar{s}\bar{s}'}(\hat H_B;t)}{n(\hat H_B)} \right\rbrack
\right\rbrack \right\}
+ O(\lambda^3) . \;
\label{dotPIss'.bis}
\end{eqnarray}
Going back to the original representation with Eq. (\ref{Pint1}) and expanding 
the two commutators, we obtain:
\begin{eqnarray}
\dot P_{ss'}(\epsilon;t)&=& -i (E_s-E_{s'}) P_{ss'}(\epsilon;t)
-\lambda^2 \sum_{\bar{s},\bar{s}'} \int^{t}_{0} d\tau \left\{ 
\frac{}{} \right. \nonumber \\ & & 
\langle s\vert \hat S \hat S(-\tau)\vert \bar{s}\rangle \langle
\bar{s}'\vert s'\rangle
{\rm Tr}_B \delta(\epsilon-\hat H_B) \hat B \hat B(-\tau)
\frac{P_{\bar{s}\bar{s}'}(\hat H_B;t)}{n(\hat H_B)}
\nonumber\\
& & -\langle s\vert \hat S \vert\bar{s}\rangle \langle \bar{s}'\vert\hat
S(-\tau)\vert s'\rangle
{\rm Tr}_B \delta(\epsilon-\hat H_B) \hat B
\frac{P_{\bar{s}\bar{s}'}(\hat H_B;t)}{n(\hat H_B)} \hat B(-\tau)\nonumber\\
& & -\langle s\vert \hat S(-\tau) \vert\bar{s}\rangle \langle
\bar{s}'\vert\hat S\vert s'\rangle
{\rm Tr}_B \delta(\epsilon-\hat H_B) \hat B(-\tau)
\frac{P_{\bar{s}\bar{s}'}(\hat H_B;t)}{n(\hat H_B)} \hat B\nonumber\\
& & \left. +
\langle s \vert \bar{s}\rangle \langle \bar{s}'\vert \hat S(-\tau) \hat
S\vert s'\rangle
{\rm Tr}_B \delta(\epsilon-\hat H_B) \frac{P_{\bar{s}\bar{s}'}(\hat
H_B;t)}{n(\hat H_B)} \hat B(-\tau) \hat B \right\} + O(\lambda^3) \; .
\label{dotPss'.bis}
\end{eqnarray}
In order to evaluate the four last terms, we notice that, for a
quasicontinuous energy spectrum, we can
write
\begin{eqnarray}
\textrm{Tr}_{B} \delta(\epsilon-\hat{H}_B) \hat{B}
\delta(\epsilon'-\hat{H}_B) \hat{B}
&=& \sum_{b,b'} \delta(\epsilon-E_b) \delta(\epsilon'-E_{b'})
\vert \langle b \vert \hat B\vert b'\rangle \vert^2 \nonumber \\
&=& n(\epsilon) n(\epsilon') F(\epsilon,\epsilon') \; , \label{fctcorrmicpauli}
\end{eqnarray}
where the function $F(\epsilon,\epsilon')$ stands for
\begin{equation}
F(\epsilon,\epsilon') \equiv ``\vert \langle \epsilon \vert \hat B \vert
\epsilon' \rangle \vert^2" \; ,
\label{F}
\end{equation}
where $\vert \epsilon \rangle$ denotes the eigenstate $\vert b \rangle$ of
the environment Hamiltonian
$\hat H_B$ corresponding to the energy eigenvalue $E_b=\epsilon$.  Equation
(\ref{F}) supposes some
smoothening of the squares $\vert \langle \epsilon \vert \hat B \vert
\epsilon' \rangle \vert^2$ of the
matrix elements of $\hat B$ over a dense spectrum of eigenvalues around the
energies $\epsilon$ and $\epsilon'$.
The function (\ref{F}) has the symmetry
\begin{equation}
F(\epsilon,\epsilon') = F(\epsilon',\epsilon) \; .
\end{equation}

With the definition (\ref{F}) and the identity
\begin{equation}
\int d\epsilon' \delta( \epsilon'-\hat H_B) = \hat I \; ,
\end{equation}
we can now write that
\begin{eqnarray}
{\rm Tr}_B \delta(\epsilon-\hat H_B) \hat B \hat B(-\tau)
\frac{P_{\bar{s}\bar{s}'}(\hat H_B;t)}{n(\hat H_B)}
&=& P_{\bar{s}\bar{s}'}(\epsilon;t) \int d\epsilon'  n(\epsilon')
F(\epsilon,\epsilon')
e^{+i(\epsilon-\epsilon')t} , \; \nonumber \\ & & \\
{\rm Tr}_B \delta(\epsilon-\hat H_B) \hat B
\frac{P_{\bar{s}\bar{s}'}(\hat H_B;t)}{n(\hat H_B)} \hat B(-\tau) &=&
n(\epsilon)  \int d\epsilon'  P_{\bar{s}\bar{s}'}(\epsilon';t)
F(\epsilon,\epsilon')
e^{+i(\epsilon-\epsilon')t} , \; \nonumber \\ & & \\
{\rm Tr}_B \delta(\epsilon-\hat H_B) \hat B(-\tau)
\frac{P_{\bar{s}\bar{s}'}(\hat H_B;t)}{n(\hat H_B)} \hat B &=& n(\epsilon)
\int d\epsilon'
P_{\bar{s}\bar{s}'}(\epsilon';t)  F(\epsilon,\epsilon')
e^{-i(\epsilon-\epsilon')t}  , \; \nonumber \\ & &
\\  {\rm Tr}_B \delta(\epsilon-\hat H_B) \frac{P_{\bar{s}\bar{s}'}(\hat
H_B;t)}{n(\hat H_B)} \hat
B(-\tau)\hat B &=& P_{\bar{s}\bar{s}'}(\epsilon;t)  \int d\epsilon'
n(\epsilon')  F(\epsilon,\epsilon') e^{-i(\epsilon-\epsilon')t}  .
\nonumber \\ & &
\end{eqnarray}
Accordingly, our quantum master equation finally takes the closed form
\begin{eqnarray}
\dot P_{ss'}(\epsilon;t)&=& -i (E_s-E_{s'}) P_{ss'}(\epsilon;t)
-\lambda^2 \sum_{\bar{s},\bar{s}'} \int d\epsilon' F(\epsilon,\epsilon')
\int^{t}_{0} d\tau \left\{ \frac{}{}
\right. \nonumber\\ & &
\langle s\vert \hat S\vert \bar{s}'\rangle \; \langle \bar{s}'\vert \hat
S\vert \bar{s} \rangle \;
P_{\bar{s}s'}(\epsilon;t) \; n(\epsilon') \;
e^{+i(\epsilon-\epsilon'+E_{\bar{s}}-E_{\bar{s}'})\tau}\nonumber\\ & &
-\langle s\vert \hat S \vert\bar{s}\rangle
\; \langle \bar{s}'\vert\hat S\vert s'\rangle \;
P_{\bar{s}\bar{s}'}(\epsilon';t) \; n(\epsilon)
 \; e^{+i(\epsilon-\epsilon'+E_{s'}-E_{\bar{s}'})\tau}\nonumber\\ & &
-\langle s\vert \hat
S\vert\bar{s}\rangle \;
\langle \bar{s}'\vert\hat S\vert s'\rangle  \;
P_{\bar{s}\bar{s}'}(\epsilon';t) \; n(\epsilon) \;
e^{-i(\epsilon-\epsilon'+E_{s}-E_{\bar{s}})\tau}\nonumber\\ & & \left. +
\langle \bar{s}' \vert \hat S \vert \bar{s}\rangle \; \langle \bar{s}\vert
\hat S\vert s'\rangle \;
P_{s\bar{s}'}(\epsilon;t) \; n(\epsilon') \;
e^{-i(\epsilon-\epsilon'+E_{\bar{s}'}-E_{\bar{s}})\tau}
\right\} + \mathcal{O}(\lambda^3) \; .
\label{our.nonMarkov.master.eq}
\end{eqnarray}
Equation (\ref{our.nonMarkov.master.eq}) determines the time
evolution of the distribution functions $P_{ss'}(\epsilon;t)$ describing
the populations and quantum coherences of
a system influencing its environment and is the central result of this
paper. It is a non-Markovian equation
because of the presence of the time integral in the right-hand side.\\

In Eq. (\ref{our.nonMarkov.master.eq}), the function $n(\epsilon) F(\epsilon,\epsilon')$
determines the properties of the coupling to the environment and, in particular, the
time scale of the environment. If this time scale is supposed to be shorter then the 
system time scales $\{ \frac{2 \pi}{E_s-E_{s'}} \}$, we can perform a
\textit{Markovian approximation} in Eq. (\ref{our.nonMarkov.master.eq}).
Such an approximation is justified for a process evolving on time scales larger
than the environment time scale. The Markovian approximation consists in
taking the limit where the upper bound of the time integral goes to infinity 
and using the following relations:
\begin{equation}
\int_{0}^{\infty} d\tau \; e^{\pm i \omega \tau} = \pm i \; \mathcal{P}
\frac{1}{\omega} + \pi \; \delta(\omega ) \; , \label{int0inf}
\end{equation}
where $\cal P$ denotes the principal part.

We finally obtain the Markovian version of our quantum master
equation (\ref{our.nonMarkov.master.eq}) as
\begin{eqnarray}
\dot P_{ss'}(\epsilon;t)&=& -i (E_s-E_{s'}) P_{ss'}(\epsilon;t) \nonumber\\
&-& i \lambda^2 \sum_{\bar{s},\bar{s}'} \left\{ \frac{}{}
\right.
\left[ \int d\epsilon' \; n(\epsilon') \; F(\epsilon,\epsilon')
\;  {\cal P}\frac{1}{\epsilon-\epsilon'+E_{\bar{s}}-E_{\bar{s}'}} \right]
\nonumber \\ & & \times \;
\left[ \langle s\vert \hat S\vert \bar{s}'\rangle \; \langle \bar{s}'\vert
\hat S\vert \bar{s} \rangle \;
P_{\bar{s}s'}(\epsilon;t) -
\langle \bar{s} \vert \hat S \vert \bar{s}'\rangle \; \langle \bar{s}'\vert
\hat S\vert s'\rangle \;
P_{s\bar{s}}(\epsilon;t) \right]\nonumber \\
& & -\langle s\vert \hat S \vert\bar{s}\rangle\; \langle \bar{s}'\vert\hat
S\vert s'\rangle \;
n(\epsilon) \nonumber \\ & & \times \int d\epsilon' \;
F(\epsilon,\epsilon') \;
P_{\bar{s}\bar{s}'}(\epsilon';t) \; \left[ {\cal P}\frac{1}{
\epsilon-\epsilon'+E_{s'}-E_{\bar{s}'}} - {\cal
P}\frac{1}{\epsilon-\epsilon'+E_{s}-E_{\bar{s}}}\right]\left.
\frac{}{}\right\} \nonumber \\
&-& \pi\; \lambda^2 \sum_{\bar{s},\bar{s}'} \left\{\frac{}{}
n(\epsilon+E_{\bar{s}}-E_{\bar{s}'})
\; F(\epsilon,\epsilon+E_{\bar{s}}-E_{\bar{s}'}) \right. \nonumber \\ & &
\times \left[
\langle s\vert \hat S\vert \bar{s}'\rangle \; \langle \bar{s}'\vert \hat
S\vert \bar{s} \rangle \;
P_{\bar{s}s'}(\epsilon;t) + \langle \bar{s} \vert \hat S \vert
\bar{s}'\rangle \; \langle \bar{s}'\vert \hat S\vert
s'\rangle \; P_{s\bar{s}}(\epsilon;t)\right] \nonumber \\ & &
-\langle s\vert \hat S \vert\bar{s}\rangle \; \langle \bar{s}'\vert\hat
S\vert s'\rangle \; n(\epsilon) \nonumber\\
& & \times \left[F(\epsilon,\epsilon+E_{s'}-E_{\bar{s}'}) \;
P_{\bar{s}\bar{s}'}(\epsilon+E_{s'}-E_{\bar{s}'};t) +
F(\epsilon,\epsilon+E_{s}-E_{\bar{s}}) \;
P_{\bar{s}\bar{s}'}(\epsilon+E_{s}-E_{\bar{s}};t) \right]
\left. \frac{}{} \right\} \nonumber \\
& & + O(\lambda^3) \; .
\label{ourMarkovmastereq}
\end{eqnarray}
We notice that the use of this Markovian equation may require a slippage of
initial conditions
as shown in Refs. \cite{GaspRed,Suarez}.  In Eq. (\ref{ourMarkovmastereq}),
the last terms in $\pi\lambda^2$ typically describe the relaxation to a
stationary solution. The terms in
$i\lambda^2$ modify the frequencies of oscillations and include the
so-called Lamb shifts of the zeroth-order
energy eigenvalues.  Indeed, if we consider only the evolution of the
off-diagonal matrix element
$P_{ss'}(\epsilon;t)$ by neglecting its coupling to all the other matrix
elements, we obtain the equation
\begin{equation}
\dot P_{ss'}(\epsilon;t) \simeq \left\{ - i\left[ \tilde
E_s(\epsilon)-\tilde E_{s'}(\epsilon)\right] -
\Gamma_{ss'}(\epsilon) \right\} \; P_{ss'}(\epsilon;t) \; ,
\end{equation}
with the energies modified by the Lamb shifts
\begin{equation}
\tilde E_s(\epsilon) = E_s + \lambda^2 \sum_{\bar{s}} \vert \langle s\vert
\hat S\vert
\bar{s}\rangle\vert^2 \;
\int d\epsilon' \; n(\epsilon') \; F(\epsilon,\epsilon') \;
{\cal P}\frac{1}{\epsilon-\epsilon'+E_{s}-E_{\bar{s}}} + O(\lambda^3)
\label{Lamb.shift}
\end{equation}
and the damping rates
\begin{eqnarray}
\Gamma_{ss'}(\epsilon) &=& \ \pi\; \lambda^2 \sum_{\bar{s}(\ne s)} \left[
\vert \langle s\vert \hat S\vert \bar{s}\rangle\vert^2 \;
n(\epsilon+E_{s}-E_{\bar{s}})\; F(\epsilon,\epsilon+E_{s}-E_{\bar{s}})
\right. \nonumber \\
& & \qquad\qquad\left. + \vert \langle s'\vert \hat S\vert
\bar{s}\rangle\vert^2 \;
n(\epsilon+E_{s'}-E_{\bar{s}})\; F(\epsilon,\epsilon+E_{s'}-E_{\bar{s}})
\right] \nonumber \\
& & + \pi\; \lambda^2 \; (\langle s\vert \hat S \vert s\rangle - \langle
s'\vert\hat S\vert s'\rangle)^2 \;
n(\epsilon)\; F(\epsilon,\epsilon) + O(\lambda^3) \; ,
\label{damping.rates}
\end{eqnarray}
in agreement with the results of Ref. \cite{Tannoudji}.
We notice that the complete equations for the off-diagonal matrix elements
couple in general different energies
because of the integrals over the environment energy $\epsilon'$.

The evolution equations for the populations of the states $\vert s\rangle$
of the system can be obtained by
neglecting the contributions from the quantum coherences, i.e., by
neglecting the terms involving off-diagonal
elements of $P_{ss'}(\epsilon;t)$.  This is justified in the weak-coupling
limit as long as the coherences vanish
or are negligible in the initial conditions, i.e., $P_{ss'}(\epsilon;0)=0$
for $s\ne s'$.  Accordingly, we obtain
the following evolution equations for the populations:
\begin{eqnarray}
\dot{P}_{ss}(\epsilon;t) &\simeq&
2\pi \lambda^2 \sum_{s'} \vert\langle s \vert \hat{S} \vert s' \rangle \vert^2
\; F(\epsilon,\epsilon+E_{s}-E_{s'}) \nonumber \\
& & \times \; \left[ \, n(\epsilon) \; P_{s's'}(\epsilon+E_{s}-E_{s'};t) -
n(\epsilon+E_{s}-E_{s'})
\; P_{ss}(\epsilon;t) \right]\nonumber . \\ \label{origpaulipop}
\end{eqnarray}
This equation is a kind of Pauli equation established with the Fermi golden
rule
and the conversation of energy in the transitions.  Indeed, if a transition
happens from a state in which the
energy of the system is $E_s$ and the one of the environment $\epsilon$ to
a state in which the system has energy
$E_{s'}$, the final energy of the environment should be
$\epsilon'=\epsilon+E_s-E_{s'}$, which is well expressed
by Eq. (\ref{origpaulipop}).  Nevertheless, Eq. (\ref{origpaulipop}) rules
the populations of the states $s$ of the system with the extra information
given by
the distribution over the environment energy $\epsilon$, which is not a
feature of
the standard Pauli equation and which turns out to be of importance
to understand the relaxation inside a nanoscopic isolated system.

Our Markovian master equation (\ref{ourMarkovmastereq}) is more general
than an equation for the
populations because it also describes the time evolution of the
distributions of the quantum coherences over the
energy of the environment.

%%%%%%%%%%%%%%%%%%%%%%%%%%%%%%%%%%%%%%%%%%%%%%%%%%%%%%%%%%%%%%%%%%%%%%%%%%%%%%%%
%
\subsection{Comparaison with the Redfield master equation}

We now discuss the conceptual differences between our quantum master
equation and another one known as the Redfield master equation.  This
equation is
well known in the context of nuclear magnetic resonance (NMR) where it
describes the time evolution of nuclear spins interacting with their
environment.

The Redfield master equation describes the time evolution of the system
density matrix
obtained tracing out from the total density matrix the degrees of
freedom of the environment
\begin{equation}
\hat{\rho}_S(t)= \textrm{Tr}_{B} \hat{\rho}(t) \; .
\label{rhosreddef}
\end{equation}
The Redfield equation is derived by using the closure approximation that
the total density matrix keeps the form
\begin{equation}
\hat{\rho}(t)=\hat{\rho}_S(t)\otimes\hat{\rho}_B 
\label{rhotevol}
\end{equation}
during the whole time evolution, where $\hat{\rho}_B$ does not depend on time.
The Redfield master equation is derived in the weak-coupling limit by a
method similar to the one of the previous section to get
\begin{eqnarray}
\dot{\hat{\rho}}_{S}(t)&=&-i \lbrack \hat{H}_S,\hat{\rho}_{S}(t) \rbrack
\nonumber \\
& &- \lambda^2 \hat{S} \int^{t}_{0} d\tau \alpha(\tau) \hat{S}(-\tau)
\hat{\rho}_S(t) \nonumber \\
& &+ \lambda^2 \hat{S} \hat{\rho}_S(t)
\int^{t}_{0} d\tau \alpha^{\ast}(\tau) \hat{S}(-\tau)  \nonumber \\
& &+ \lambda^2 \int^{t}_{0} d\tau \alpha(\tau) \hat{S}(-\tau)
\hat{\rho}_S(t) \hat{S} \nonumber \\
& &- \lambda^2 \hat{\rho}_S(t)
\int^{t}_{0} d\tau \alpha^{\ast}(\tau) \hat{S}(-\tau)  \hat{S}
+ O(\lambda^3) \; ,
\label{rhoSpertgendevv}
\end{eqnarray}
with the correlation function of the environment operators
\begin{equation}
\alpha(t) = \langle \hat{B}(t) \hat{B}(0) \rangle=\textrm{Tr}_{B}
\hat{\rho}_B \hat{B}(t) \hat{B}(0) \; .
\label{defcorrel}
\end{equation}
Equation (\ref{rhoSpertgendevv}) is a \textit{non-Markovian Redfield
equation}. The non-Markovianity
comes from the fact that the integrals over expressions containing the
correlation function depend on time.
The density matrix of the system can be represented in the basis of the
system eigenstates as
\begin{equation}
\eta_{ss'}(t) \equiv \langle s\vert\hat{\rho}_{S}(t)\vert s'\rangle \; .
\label{eta}
\end{equation}
In this representation, the non-Markovian
Redfield equation has the following form:
\begin{eqnarray}
\dot\eta_{ss'}(t) &=& -i (E_s-E_{s'}) \;
\eta_{ss'}(t)
-\lambda^2 \sum_{\bar{s},\bar{s}'} \int d\epsilon' n(\epsilon')
F(\epsilon,\epsilon') \int^{t}_{0} d\tau \left\{ \frac{}{}
\right. \nonumber\\ & &
\ \langle s\vert \hat S\vert \bar{s}'\rangle \; \langle \bar{s}'\vert \hat
S\vert \bar{s} \rangle \;
\eta_{\bar{s}s'}(t) \;
e^{+i(\epsilon-\epsilon'+E_{\bar{s}}-E_{\bar{s}'})\tau}\nonumber\\ & &
-\langle s\vert \hat S \vert\bar{s}\rangle
\; \langle \bar{s}'\vert\hat S\vert s'\rangle \;
\eta_{\bar{s}\bar{s}'}(t)
\; e^{+i(\epsilon-\epsilon'+E_{\bar{s}}-E_{s})\tau}\nonumber\\ & &
-\langle s\vert \hat S\vert\bar{s}\rangle \;
\langle \bar{s}'\vert\hat S\vert s'\rangle  \;  \eta_{\bar{s}\bar{s}'}(t) \;
e^{-i(\epsilon-\epsilon'+E_{\bar{s}'}-E_{s'})\tau}\nonumber\\ & & \left. +
\langle \bar{s}' \vert \hat S \vert \bar{s}\rangle \; \langle \bar{s}\vert
\hat S\vert s'\rangle \;
\eta_{s\bar{s}'}(t) \;
e^{-i(\epsilon-\epsilon'+E_{\bar{s}'}-E_{\bar{s}})\tau}
\right\} \nonumber \\
& & + O(\lambda^3) \; .
\label{Redfield.nonMarkov.master.eq}
\end{eqnarray}

If the environment is large enough, the correlation function in Eq.
(\ref{rhoSpertgendevv}) goes to zero after a
certain time. This time, called the environment correlation time $\tau_{\rm
corr}$, determines the time scale of
the environment dynamics. If we perform the \textit{Markovian
approximation} that consists of putting the upper
bound of the time integral in the non-Markovian Redfield equation to
infinity, one gets the standard Redfield
equation. We notice that, in doing so, the time evolution may be spoiled on
a time scale of order $\tau_{\rm
corr}$ unless some use is made of some slipped initial conditions
\cite{GaspRed,Suarez}.  Performing this Markovian
approximation, one gets the standard (Markovian) \textit{Redfield
equation} given by Eq. (\ref{rhoSpertgendevv}) with $\int_{0}^{t}$
replaced by $\int_{0}^{\infty}$.
%\begin{eqnarray}
%\dot{\hat{\rho}}_{S}(t)&=&-i \lbrack \hat{H}_S,\hat{\rho}_{S}(t) \rbrack
%\nonumber \\
%& &- \lambda^2 \hat{S}  \int^{\infty}_{0} d\tau \alpha(\tau) \hat{S}(-\tau)
%\hat{\rho}_S(t) \nonumber \\
%& &+ \lambda^2 \hat{S} \hat{\rho}_S(t)
%\int^{\infty}_{0} d\tau \alpha^{\ast}(\tau) \hat{S}(-\tau)   \nonumber \\
%& &+ \lambda^2  \int^{\infty}_{0} d\tau \alpha(\tau) \hat{S}(-\tau)
%\hat{\rho}_S(t) \hat{S} \nonumber \\
%& &- \lambda^2 \hat{\rho}_S(t)
%\int^{\infty}_{0} d\tau \alpha^{\ast}(\tau) \hat{S}(-\tau) \hat{S}
%+ O(\lambda^3).
%\label{markrhoSpertgendevv}
%\end{eqnarray}
As shown in Refs. \cite{GaspRed,Suarez}, the use of this Redfield Markovian
equation needs to be supplemented by a slippage of initial conditions.

In order to compare the Redfield equation with our master equation derived
in the previous section, we consider the
case where the environment is initially in the \textit{microcanonical} state:
\begin{equation}
\hat{\rho}_B=\frac{\delta(\epsilon-\hat{H}_B)}{\textrm{Tr}_{B}
\delta(\epsilon-\hat{H}_B)}
=\frac{\delta(\epsilon-\hat{H}_B)}{n(\epsilon)} \; .
\label{matdensmicro}
\end{equation}
Having chosen the microcanonical density matrix (\ref{matdensmicro}) for the
environment, the correlation function (\ref{defcorrel}) takes the form
\begin{eqnarray}
\alpha(\tau,\epsilon) &=& \frac{1}{n(\epsilon)}  \;
{\rm Tr}_{B} \delta(\epsilon-\hat{H}_B) \hat{B}(\tau) \hat{B} \nonumber \\
&=& \frac{1}{n(\epsilon)} \; \int d\epsilon' \; {\rm Tr}_{B}
\delta(\epsilon-\hat{H}_B) \hat{B}(\tau)
\delta(\epsilon'-\hat{H}_B) \hat{B} \nonumber \\
&=& \int d\epsilon' \; n(\epsilon') \;
F(\epsilon,\epsilon') \; e^{i(\epsilon-\epsilon')\tau} \; .
\label{fctcorrmicred}
\end{eqnarray}

In the basis of the system eigenstates, the Redfield equation takes the form
\begin{eqnarray}
\dot\eta_{ss'}(t) &=& -i (E_{s}-E_{s'}) \; \eta_{ss'}(t) \nonumber\\
&-& i \lambda^2 \sum_{\bar{s},\bar{s}'} \left\{ \frac{}{}
\right.
\left[ \int d\epsilon' \; n(\epsilon') \; F(\epsilon,\epsilon')
\;  {\cal P}\frac{1}{\epsilon-\epsilon'+E_{\bar{s}}-E_{\bar{s}'}} \right]
\nonumber \\ & & \times \;
\left[ \langle s\vert \hat S\vert \bar{s}'\rangle \; \langle \bar{s}'\vert
\hat S\vert \bar{s} \rangle \;
\eta_{\bar{s}s'}(t) -
\langle \bar{s} \vert \hat S \vert \bar{s}'\rangle \; \langle \bar{s}'\vert
\hat S\vert s'\rangle \;
\eta_{s\bar{s}}(t) \right]\nonumber \\
& & + \int d\epsilon' \; n(\epsilon') \; F(\epsilon,\epsilon') \; \left[
{\cal P}\frac{1}{
\epsilon-\epsilon'+E_{\bar{s}'}-E_{s'}} - {\cal
P}\frac{1}{\epsilon-\epsilon'+E_{\bar{s}}-E_{s}}\right] \nonumber \\
& & \times \; \langle s\vert \hat S \vert\bar{s}\rangle\; \langle
\bar{s}'\vert\hat S\vert s'\rangle \;
\eta_{\bar{s}\bar{s}'}(t)  \left. \frac{}{}\right\} \nonumber \\
&-& \pi\; \lambda^2 \sum_{\bar{s},\bar{s}'} \left\{\frac{}{}
n(\epsilon+E_{\bar{s}}-E_{\bar{s}'})
\; F(\epsilon,\epsilon+E_{\bar{s}}-E_{\bar{s}'}) \right. \nonumber \\ & &
\times \left[
\langle s\vert \hat S\vert \bar{s}'\rangle \; \langle \bar{s}'\vert \hat
S\vert \bar{s} \rangle \;
\eta_{\bar{s}s'}(t)
+ \langle \bar{s} \vert \hat S \vert \bar{s}'\rangle \; \langle
\bar{s}'\vert \hat S\vert
s'\rangle \; \eta_{s\bar{s}}(t)\right] \nonumber \\ & &
-\left[ n(\epsilon+E_{\bar{s}'}-E_{s'}) \;
F(\epsilon,\epsilon+E_{\bar{s}'}-E_{s'})
+ n(\epsilon+E_{\bar{s}}-E_{s}) \; F(\epsilon,\epsilon+E_{\bar{s}}-E_{s})
\right] \nonumber \\ & &
\times \; \langle s\vert \hat S \vert\bar{s}\rangle \; \langle
\bar{s}'\vert\hat S\vert s'\rangle \;
\eta_{\bar{s}\bar{s}'}(t) \left. \frac{}{} \right\} \nonumber \\
& & + O(\lambda^3) \; .
\label{Redfield.Markov.master.eq}
\end{eqnarray}

The off-diagonal elements of the system density matrix individually obey
the equations
\begin{equation}
\dot\eta_{ss'}(t) \simeq \left\{ - i\left[ \tilde E_s(\epsilon)-\tilde
E_{s'}(\epsilon)\right] -
\Gamma_{ss'}(\epsilon) \right\} \; \eta_{ss'}(t) \; ,
\end{equation}
with the same Lamb shifts (\ref{Lamb.shift}) and damping rates
(\ref{damping.rates}) as in our master equation
and as expected from Ref. \cite{Tannoudji}.
There is no difference between our quantum master equation and the Redfield
one at this stage.

On the other hand, the Redfield equation predicts an evolution of the
populations ruled by the following equation
obtained by neglecting all the contributions coming from the coherences in
Eq. (\ref{Redfield.Markov.master.eq}):
\begin{eqnarray}
\dot\eta_{ss}(t) &=&
2\pi \lambda^2 \sum_{s'} \vert\langle s \vert \hat{S} \vert s' \rangle
\vert^2 \nonumber\\
& & \quad \times \; \left[
F(\epsilon,\epsilon+E_{s'}-E_{s}) \;  n(\epsilon+E_{s'}-E_{s}) \;
\eta_{s's'}(t) \right. \nonumber \\ & & \quad -
F(\epsilon,\epsilon+E_{s}-E_{s'}) \;  n(\epsilon+E_{s}-E_{s'}) \;
\eta_{ss}(t) \left.\right] \; .
\label{cohentanpop}
\end{eqnarray}
This equation is the same as the master equation for the populations derived
by Cohen-Tannoudji and co-workers in Ref. \cite{Tannoudji}.

We notice that important differences now exists between the population
equation (\ref{cohentanpop}) obtained from
the Redfield equation and the other population equation (\ref{origpaulipop})
obtained from our master equation. Both equations describe the
evolution of the populations as a random walk process in the spectrum. However,
these processes are significantly different for Eqs. (\ref{origpaulipop})
and (\ref{cohentanpop}). Let us focus
on the evolution of the probability to be on a system state corresponding
to the system energy $E_s$. In both
equations we see that for the loss contributions to the
evolution coming from the jumps from an energy $E_s$ to an
energy $E_{s'}$, the density of states of the environment is
modified by the energy $E_s-E_{s'}$. This is consistent with the
Fermi golden rule applied to the total system and, thus, keeps
the total energy constant. We care now on the gain
contributions to the evolution. In our equation (\ref{origpaulipop}), we
see that
for these contributions due to jumps from the system
energy $E_{s'}$ to $E_{s}$, the density of states of the
environment is modified by the energy $E_{s'}-E_s$. This is also
consistent with the Fermi golden rule applied to the total system
and, thus, keeps the total energy constant. However, for the Redfield
equation we see that for the jumps from an energy $E_{s'}$ to an
energy $E_{s}$, the density of states of the environment is not
modified by the energy $E_{s'}-E_s$, which is not
consistent with the Fermi golden rule applied to the total system
and does not keep the total energy constant.

\begin{figure}[P]
\centering
\rotatebox{0}{\scalebox{0.8}{\includegraphics{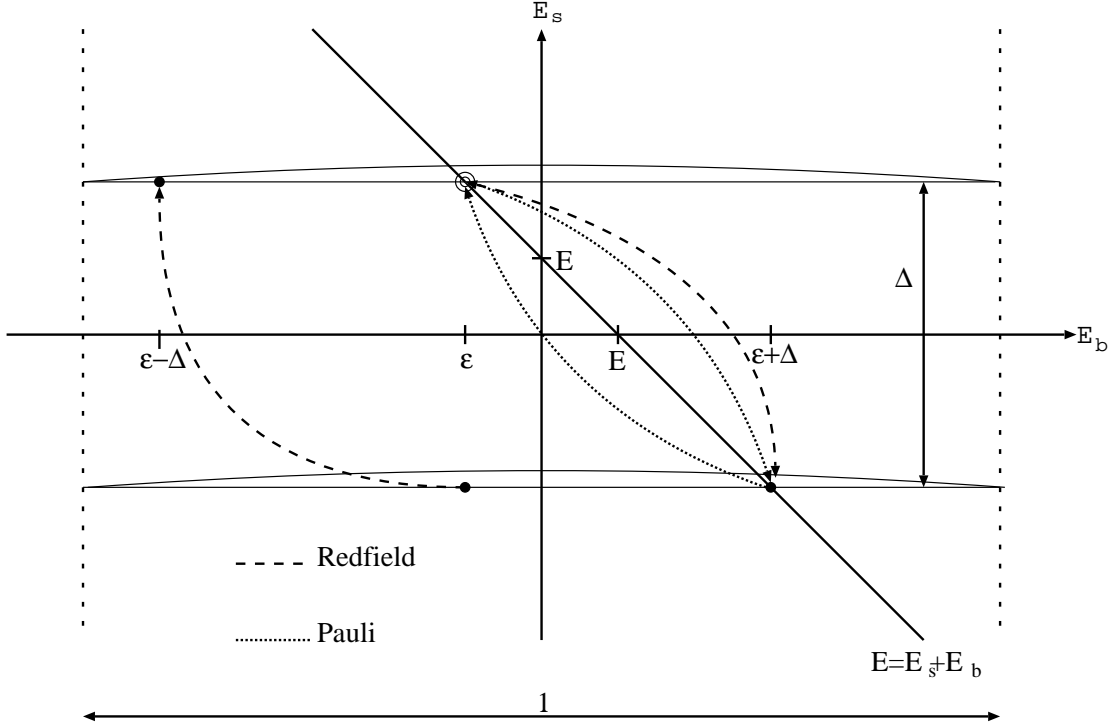}}} \\
\caption{Schematic representation of the energy exchanges described,
respectively, by the Redfield and by our master equation
in the Markovian limit for a two-level system model, in the
plane of the system energy $E_S$ versus the environment energy $E_B$.
The energy splitting between the two levels of the system is denoted by
$\Delta$. The energy spectrum of the system is discrete (two levels) while
the one of the environment is a quasicontinuum represented
by the density of states given by the Wigner semicircle law (\ref{semi-circ})
of width equal to unity. The total energy of
the system is given by $E=E_S+E_B$, which corresponds to the diagonal line.
The initial condition is denoted by two empty superposed circles.
We see that transitions preserving
the total energy have to occur along the diagonal line $E=E_S+E_B$.
Doing this, they satisfy the Fermi golden rule for the total system.
One can see that only our master equation satisfies this condition (dotted
transition lines). The Redfield
equation describes transitions that occur along a vertical line at
constant environment energy and is therefore wrong (dashed transition lines).
\label{sautredpaulispingoe}}
\end{figure}

\begin{figure}[P]
\centering
\rotatebox{0}{\scalebox{0.8}{\includegraphics{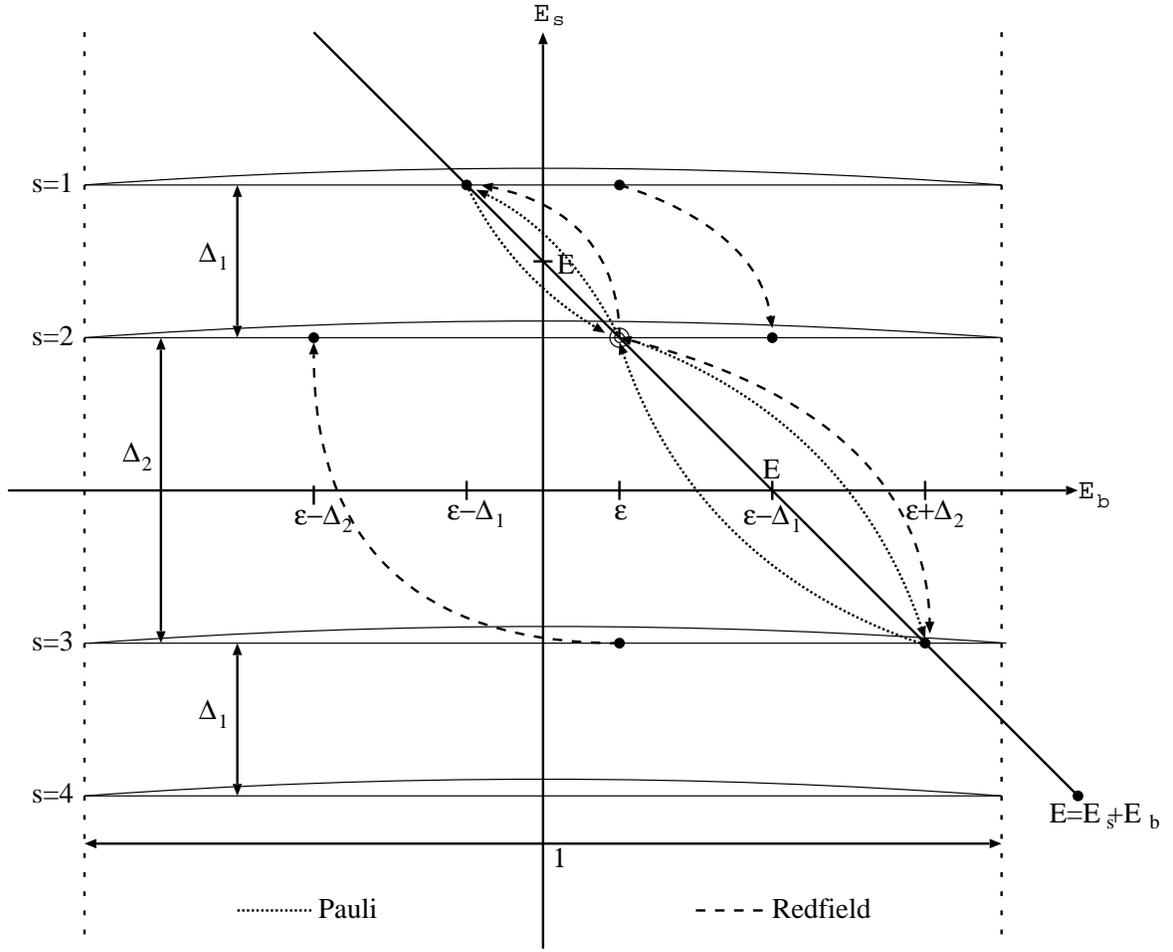}}} \\
\caption{Generalization of the previous Fig. \ref{sautredpaulispingoe} for
the case where the system
has more than two levels (here four levels). One can see that the system
levels (horizontal lines) that do not
intersect the total energy diagonal line $E=E_S+E_B$ within the environment
energy spectrum delimited
by the sparse-dotted vertical lines do not participate in the dynamics.
\label{sautredpauligen}}
\end{figure}

One can represent, for the Markovian case, the transitions described by
Eq. (\ref{origpaulipop}) and (\ref{cohentanpop}) in a plane of the
system energy versus the environment energy.  In Figs.
\ref{sautredpaulispingoe} and
\ref{sautredpauligen} we have depicted the energy exchanges
described, respectively, by the Redfield equation and our equation in
the Markovian limit for two different systems.
Transitions between the system and the environment have to
preserve the energy of the total system according to the Fermi golden rule
and have therefore to occur along diagonal lines of the plane. One can see in
Figs. \ref{sautredpaulispingoe} and \ref{sautredpauligen}
that only our equation satisfies this condition. The Redfield
equation describes transitions that occur along a vertical line at
constant environment energy and is therefore not consistent with
energy conservation in the total system.  This is acceptable if the
environment is sufficiently large and has an arbitarily large energy.
However, this is inadequate if the total energy of the system and the
environment is
finite as in nanoscopic systems, in which case our master should replace
the Redfield equation.

We can summarize as follows the differences between our quantum master equation
(\ref{ourMarkovmastereq}) and the Redfield equation
(\ref{Redfield.Markov.master.eq}).
The derivation of both equations is based on the perturbative
expansion of the total density matrix, but a specific form is imposed in
each equation
to the total density matrix [see Eqs. (\ref{formegenrhopauli})
and (\ref{rhotevol})]. The consequence of this choice can be
seen on the reduced density matrix of the system. In the Redfield
theory, we have
\begin{equation}
\hat{\rho}_S(t)=\sum_{s,s'} \vert s\rangle \; \eta_{ss'}(t) \; \langle s'
\vert \;,
\end{equation}
while, for our master equation, using Eq. (\ref{rhospaulimic}) we have
\begin{equation}
\hat{\rho}_S(t)=\sum_{s,s'} \int d\epsilon \; \vert s\rangle \;
P_{ss'}(\epsilon;t) \; \langle s'
\vert \; .
\end{equation}
The system density matrix is related to the distribution functions according to
\begin{equation}
\langle s \vert \hat{\rho}_S(t) \vert s'\rangle =\eta_{ss'}(t)= \int
d\epsilon \; P_{ss'}(\epsilon;t) \; .
\label{rho_S}
\end{equation}
We see that, in our master equation, the matrix elements of the
system density matrix are decomposed on the energy of the environment.
This is not the case for the Redfield equation. The decomposition allows us to
correlate the states of the system with the states of the
environment. This is the main point of our master equation. The density
matrix adopted for the Redfield equation
cannot describe such correlations. In the Redfield equation, during the
evolution,
the environment is always in the same state while, in our master
equation, the state of the environment is determined by the state
of the system.  As a consequence, we obtain a description which is consistent
with energy conservation thanks to our master equation.

%%%%%%%%%%%%%%%%%%%%%%%%%%%%%%%%%%%%%%%%%%%%%%%%%%%%%%%%%%%%%%%%%%%%%%%%%%%%%%%%
%
%%%%%%%%%%%%%%%%%%%%%%%%%%%%%%%%%%%%%%%%%%%%%%%%%%%%%%%%%%%%%%%%%%%%%%%%%%%%%%%%
%
\section{Application to the spin-environment model \label{spin-boson}}

In this section we consider a specific class of two-level systems
interacting with an environment. The two-level system may be supposed to be
a spin.
An example is the spin-boson model in which the environment
is a set of harmonic oscillators behaving as phonons \cite{Legget}.

The Hamiltonian of the spin-environment model we consider here is the
following:
\begin{eqnarray}
\hat{H}_{\rm tot}=\frac{\Delta}{2} \hat{\sigma}_{z}+\hat{H}_{B}+\lambda
\hat{\sigma}_x \hat{B} \; .
\label{hamiltonienspinboson}
\end{eqnarray}
The eigenvalue equation of the system is
\begin{eqnarray}
\hat H_S \vert s \rangle = \frac{\Delta}{2} \hat{\sigma}_{z} \vert s
\rangle = s \frac{\Delta}{2} \vert s \rangle
\; ,
\label{etproprespin}
\end{eqnarray}
where $s=\pm 1$. Like in Sec. \ref{genpauleq}, we first
derive our master equation and then the Redfield equation in order to
compare both equations.

\subsection{Using our master equation}

Let us now apply our master equation to the spin-environment model.

In our theory and for a two-level system, the total density
matrix becomes
\begin{eqnarray}
\hat{\rho}(t)&=&\frac{1}{n(\hat H_B)} \left[ P_{++}(\hat{H}_B;t) \vert +
\rangle \langle + \vert
+ P_{+-}(\hat{H}_B;t)  \vert + \rangle \langle - \vert \right.\nonumber \\
& & \left. \qquad\qquad + P_{-+}(\hat{H}_B;t)  \vert - \rangle \langle + \vert
+ P_{--}(\hat{H}_B;t) \vert - \rangle \langle - \vert \right] \;
.\label{rhosspinbos}
\end{eqnarray}
For the spin-environment model, our non-Markovian master equation
(\ref{our.nonMarkov.master.eq}) is given by
\begin{eqnarray}
\dot{P}_{++}(\epsilon;t) &=& - \lambda^2 P_{++}(\epsilon;t) \int d\epsilon'
F(\epsilon,\epsilon')
n(\epsilon') \int_{0}^{t} d\tau
\left[ e^{i (\epsilon-\epsilon'+\Delta) \tau}+e^{-i
(\epsilon-\epsilon'+\Delta) \tau}\right] \nonumber \\
& & + \lambda^2 n(\epsilon) \int d\epsilon' F(\epsilon,\epsilon')
P_{--}(\epsilon';t) \int_{0}^{t} d\tau
\left[e^{i (\epsilon-\epsilon'+\Delta) \tau}+e^{-i
(\epsilon-\epsilon'+\Delta) \tau}\right] \; ,\nonumber \\
\label{paulispinbosNM11}
\end{eqnarray}
\begin{eqnarray}
\dot{P}_{--}(\epsilon;t) &=& - \lambda^2 P_{--}(\epsilon;t) \int d\epsilon'
F(\epsilon,\epsilon') n(\epsilon')
\int_{0}^{t} d\tau \left[e^{i (\epsilon-\epsilon'-\Delta) \tau}+e^{-i
(\epsilon-\epsilon'-\Delta)\tau}\right]
\nonumber \\ & & + \lambda^2 n(\epsilon) \int d\epsilon'
F(\epsilon,\epsilon') P_{++}(\epsilon';t)
\int_{0}^{t} d\tau \left[ e^{i (\epsilon-\epsilon'-\Delta) \tau}+e^{-i
(\epsilon-\epsilon'-\Delta)
\tau}\right] \; ,\nonumber \\
\label{paulispinbosNM-1-1}
\end{eqnarray}
\begin{eqnarray}
\dot{P}_{+-}(\epsilon;t) &=& -i \Delta \; P_{+-}(\epsilon;t) \nonumber \\
& & - \lambda^2 P_{+-}(\epsilon;t) \int d\epsilon' F(\epsilon,\epsilon')
n(\epsilon')
\int_{0}^{t} d\tau \left[ e^{i (\epsilon-\epsilon'+\Delta) \tau}+e^{-i
(\epsilon-\epsilon'-\Delta) \tau}\right]
\nonumber \\ & & + \lambda^2 n(\epsilon) \int d\epsilon'
F(\epsilon,\epsilon') P_{-+}(\epsilon';t)
\int_{0}^{t} d\tau \left[ e^{i (\epsilon-\epsilon'-\Delta) \tau}+e^{-i
(\epsilon-\epsilon'+\Delta)
\tau}\right] \; .\nonumber \\
\label{paulispinbosNM1-1}
\end{eqnarray}
and a further equation for $\dot{P}_{-+}(\epsilon;t)$ given by the complex 
conjugate of Eq. (\ref{paulispinbosNM1-1}).\\

We observe that the diagonal and off-diagonal elements of
$P_{ss'}(\epsilon;t)$ obey decoupled equations in the case of 
the spin-environment model.  
Therefore, the time evolution of the populations
is independent of the time evolution of the quantum coherences.
We now perform the \textit{Markovian approximation} that consists of
putting the upper bound of the time integral
to infinity. Using Eq. (\ref{int0inf}), we find
\begin{equation}
\dot{P}_{++}(\epsilon;t) = 2 \pi \lambda^2 F(\epsilon,\epsilon+\Delta)
\left[ n(\epsilon) P_{--}(\epsilon+\Delta;t)
-n(\epsilon+\Delta) P_{++}(\epsilon;t)\right] \; ,
\label{paulispinbosM11}
\end{equation}
\begin{equation}
\dot{P}_{--}(\epsilon;t) = 2 \pi \lambda^2 F(\epsilon,\epsilon-\Delta)
\left[ n(\epsilon) P_{++}(\epsilon-\Delta;t)
-n(\epsilon-\Delta) P_{--}(\epsilon;t)\right] \; ,
\label{paulispinbosM-1-1}
\end{equation}
\begin{eqnarray}
\dot{P}_{+-}(\epsilon;t) &=& -i \Delta \; P_{+-}(\epsilon;t) \nonumber \\
& & +i \lambda^2 \int d\epsilon' F(\epsilon,\epsilon')
{\cal P}\frac{2\Delta}{(\epsilon-\epsilon')^2-\Delta^2}
\left[ n(\epsilon') {P}_{+-}(\epsilon;t) + n(\epsilon)
{P}_{-+}(\epsilon';t)\right] \nonumber\\
& & - \pi \lambda^2 \left[ n(\epsilon+\Delta) F(\epsilon,\epsilon+\Delta)
+ n(\epsilon-\Delta) F(\epsilon,\epsilon-\Delta)\right] \;
P_{+-}(\epsilon;t) \nonumber \\
& & + \pi \lambda^2 \; n(\epsilon) \left[ F(\epsilon,\epsilon+\Delta)
P_{-+}(\epsilon+\Delta;t)
+ F(\epsilon,\epsilon-\Delta) P_{-+}(\epsilon-\Delta;t) \right] \;
. \nonumber \\
\label{paulispinbosM1-1}
\end{eqnarray}
We notice that during the time evolution of the populations, the following
quantity remains a constant of motion:
\begin{equation}
P(\epsilon;t) \equiv P_{++}(\epsilon;t) +
P_{--}(\epsilon+\Delta;t) = P(\epsilon;0) \; .
\label{cst}
\end{equation}
Accordingly, the difference of the populations defined as
\begin{equation}
Z(\epsilon;t) \equiv P_{++}(\epsilon;t) - P_{--}(\epsilon+\Delta;t) 
\label{def.Z}
\end{equation}
obeys the differential equation
\begin{eqnarray}
\dot Z(\epsilon;t) &=& 2\pi \lambda^2 \left[ n(\epsilon)-
n(\epsilon+\Delta)\right] F(\epsilon,\epsilon+\Delta) 
P(\epsilon;0) \nonumber\\
& & - 2\pi \lambda^2 \left[ n(\epsilon) +
n(\epsilon+\Delta) \right] F(\epsilon,\epsilon+\Delta)
Z(\epsilon;t)\; ,
\label{eq.Z}
\end{eqnarray}
the solution of which is given by
\begin{equation}
Z(\epsilon;t) = Z(\epsilon;\infty) + \left[ Z(\epsilon;0)
-Z(\epsilon;\infty)\right]\; e^{-\gamma_{\rm Pauli} t}
\end{equation}
with the asymptotic equilibrium value
\begin{equation}
Z(\epsilon;\infty)  = \frac{n(\epsilon)-
n(\epsilon+\Delta)}{n(\epsilon)+
n(\epsilon+\Delta)}\; P(\epsilon;0) 
\end{equation}
and the relaxation rate
\begin{equation}
\gamma_{\rm Pauli} = 2\pi \lambda^2 \left[ n(\epsilon) +
n(\epsilon+\Delta) \right] F(\epsilon,\epsilon+\Delta) \; .
\end{equation}
Therefore, the populations relax to their asymptotic equilibrium values
for each pair of energies $\epsilon$ and $\epsilon+\Delta$ of the environment,
keeping constant the initial distribution of the quantity $P(\epsilon;0)$.

The time evolution of the distribution functions $P_{\pm\mp}(\epsilon;t)$
of the quantum coherences is more complicated because there is now a coupling
between a continuum of values of the environment energy instead of only two
values.
Accordingly, the distributions of quantum coherence is ruled by a couple of two
integro-differential equations, instead of an ordinary differential equation.

\subsection{Using the Redfield equation}

For the spin-environment model, the non-Markovian and Markovian Redfield
equations can be derived from Eqs.
(\ref{Redfield.nonMarkov.master.eq}) and (\ref{Redfield.Markov.master.eq}).

Using Eq. (\ref{Redfield.nonMarkov.master.eq}), the non-Markovian Redfield
equations here write
\begin{eqnarray}
\dot\eta_{++}(t)&=&
-\lambda^2 \; \eta_{++}(t)
\int d\epsilon' n(\epsilon') F(\epsilon,\epsilon')
\int_{0}^{t} d\tau \left[
e^{i(\epsilon-\epsilon'+\Delta)\tau}+e^{-i(\epsilon-\epsilon'+\Delta)\tau}\right
]
\nonumber \\ & &+\lambda^2 \;  \eta_{--}(t)
\int d\epsilon' n(\epsilon') F(\epsilon,\epsilon')
\int_{0}^{t} d\tau \left[
e^{i(\epsilon-\epsilon'-\Delta)\tau}+e^{-i(\epsilon-\epsilon'-\Delta)\tau}\right
]
\; ,\nonumber \\ \label{redspinboscorrNM11}
\end{eqnarray}
\begin{eqnarray}
\dot\eta_{--}(t)&=&
-\lambda^2 \;  \eta_{--}(t)
\int d\epsilon' n(\epsilon') F(\epsilon,\epsilon')
\int_{0}^{t} d\tau \left[
e^{i(\epsilon-\epsilon'-\Delta)\tau}+e^{-i(\epsilon-\epsilon'-\Delta)\tau}\right
]
 \nonumber \\
& &+\lambda^2 \;  \eta_{++}(t)
\int d\epsilon' n(\epsilon') F(\epsilon,\epsilon')
\int_{0}^{t} d\tau
\left[e^{i(\epsilon-\epsilon'+\Delta)\tau}+e^{-i(\epsilon-\epsilon'+\Delta)\tau}
\right]
\; ,\nonumber \\ \label{redspinboscorrNM-1-1}
\end{eqnarray}
\begin{eqnarray}
\dot\eta_{+-}(t) &=& -i \Delta \; \eta_{+-}(t) \nonumber \\
& & -\lambda^2 \;  \eta_{+-}(t)
\int d\epsilon' n(\epsilon') F(\epsilon,\epsilon')
\int_{0}^{t} d\tau
\left[e^{i(\epsilon-\epsilon'+\Delta)\tau}+e^{-i(\epsilon-\epsilon'-\Delta)\tau}
\right] \nonumber\\ & & +\lambda^2 \;
\eta_{-+}(t)
\int d\epsilon' n(\epsilon') F(\epsilon,\epsilon')
\int_{0}^{t} d\tau
\left[e^{i(\epsilon-\epsilon'-\Delta)\tau}+e^{-i(\epsilon-\epsilon'+\Delta)\tau}
\right] \; ,\nonumber \\
\label{redspinboscorrNM1-1}
\end{eqnarray}
and a further equation for $\eta_{-+}(t)$ given by the complex conjugate of 
Eq. (\ref{redspinboscorrNM1-1}).\\

Here again, there is a decoupling between the time evolutions of the
populations and of the
quantum coherences. \\

Taking the Markovian approximation by replacing $\int_0^t$ into
$\int_0^{\infty}$ and using Eq. (\ref{int0inf}), we
get the Markovian Redfield equations for the spin-environment model:
\begin{eqnarray}
\dot\eta_{++}(t) &=& 2\pi\lambda^2 \left[
n(\epsilon-\Delta) \; F(\epsilon,\epsilon-\Delta) \; \eta_{--}(t)
\right.\nonumber \\ & & \left. \qquad
- n(\epsilon+\Delta) \; F(\epsilon,\epsilon+\Delta) \; \eta_{++}(t)\right]
\; ,
\label{redspinboscorrM11}
\end{eqnarray}
\begin{eqnarray}
\dot\eta_{--}(t) &=& 2\pi\lambda^2 \left[
n(\epsilon + \Delta) \; F(\epsilon,\epsilon+\Delta) \; \eta_{++}(t)
\right.\nonumber \\ & & \left. \qquad
- n(\epsilon-\Delta) \; F(\epsilon,\epsilon-\Delta) \; \eta_{--}(t)\right] \; , 
\label{redspinboscorrM-1-1} 
\end{eqnarray}
\begin{eqnarray}
\dot\eta_{+-}(t)&
=& -i \Delta \; \eta_{+-}(t) \nonumber \\
& & +i \lambda^2 \int d\epsilon' n(\epsilon') F(\epsilon,\epsilon')
{\cal P}\frac{2\Delta}{(\epsilon-\epsilon')^2-\Delta^2}
\left[  \eta_{+-}(t)
+ \eta_{-+}(t) \right]
\nonumber\\
& & -\pi \lambda^2
\left[ n(\epsilon+\Delta) F(\epsilon,\epsilon+\Delta) + n(\epsilon-\Delta)
F(\epsilon,\epsilon-\Delta)\right]
\nonumber \\ & & \qquad\qquad\qquad\qquad\qquad\qquad\qquad\qquad \times \;
\left[ \eta_{+-}(t)
- \eta_{-+}(t) \right] \; .
\nonumber \\ \label{redspinboscorrM1-1}
\end{eqnarray}

The populations of the two-level system are controlled by the
$z$ component of
the spin defined as the difference
\begin{equation}
z_{\rm Redfield}(t) = \eta_{++}(t) -\eta_{--}(t) \; .
\label{redspinbosonMZtdef}
\end{equation}
According to the Markovian Redfield equations (\ref{redspinboscorrM11}) and
(\ref{redspinboscorrM-1-1}),
the $z$ component obeys the differential equation
\begin{eqnarray}
\dot z_{\rm Redfield} &=& 2\pi \lambda^2 \left[ n(\epsilon-\Delta)
F(\epsilon,\epsilon-\Delta)
- n(\epsilon+\Delta) F(\epsilon,\epsilon+\Delta)\right] \nonumber \\
& & - 2\pi\lambda^2 \left[ n(\epsilon-\Delta) F(\epsilon,\epsilon-\Delta)
+ n(\epsilon+\Delta) F(\epsilon,\epsilon+\Delta)\right] \; z_{\rm Redfield}
\; . \nonumber\\
\end{eqnarray}
Its solution is given by
\begin{equation}
z_{\rm Redfield}(t) = z_{\rm Redfield}(\infty) + \left[z_{\rm Redfield}(0)
- z_{\rm Redfield}(\infty) \right]
\; e^{-\gamma_{\rm Redfield} t} \; , \label{redspinbosonMZt}
\end{equation}
with the equilibrium value
\begin{equation}
z_{\rm Redfield}(\infty)=\frac{n(\epsilon-\Delta) F(\epsilon,\epsilon-\Delta)
- n(\epsilon+\Delta) F(\epsilon,\epsilon+\Delta)}{
n(\epsilon-\Delta) F(\epsilon,\epsilon-\Delta)
+ n(\epsilon+\Delta) F(\epsilon,\epsilon+\Delta)} \; ,
\label{redspinbosonMZinfini}
\end{equation}
and the relaxation rate
\begin{equation}
\gamma_{\rm Redfield}=2\pi\lambda^2 \left[ n(\epsilon-\Delta)
F(\epsilon,\epsilon-\Delta)
+ n(\epsilon+\Delta) F(\epsilon,\epsilon+\Delta)\right] \; .
\label{redspinbosonMrate}
\end{equation}
We notice that the rate predicted by the Redfield equation
coincides with the one predicted by our master equation only in the
limit $\Delta=0$.  A more important difference appears
in the asymptotic equilibrium values for the $z$ component
of the spin predicted by both equations.  These
differences find their origin in the problem of conservation
of energy with the Redfield equation, as explained above.
Comparison with numerical data will confirm this explanation in a following
section
in the case of the spin-GORM model.

%%%%%%%%%%%%%%%%%%%%%%%%%%%%%%%%%%%%%%%%%%%%%%%%%%%%%%%%%%%%%%%%%%%%%%%%%%%%%%%%
%
%%%%%%%%%%%%%%%%%%%%%%%%%%%%%%%%%%%%%%%%%%%%%%%%%%%%%%%%%%%%%%%%%%%%%%%%%%%%%%%%
%
\section{Application to the spin-GORM model \label{spin-GORM}}

In order to confront our master equation and the Redfield equation with
numerical data
and test their respective domains of validity,
we now apply our theory to a specific class
of two-level systems interacting with an environment, for which
the environment operators are Gaussian orthogonal random
matrices (GORM). We call this model the spin-GORM model
and its detailed properties will be described elsewhere
\cite{EspoGasp}.

The system is a two-level system, while the environment is
supposed to be a system with a very complex dynamics.
Here, the term complex is used in a generic way.
The complexity can come, for example, from the fact that the
corresponding classical system is chaoticlike in a quantum
billiard or for the hydrogen atom in a strong magnetic field
\cite{BGS,Guhr}. It can also come from a large number of coupling
between states in an interacting many-body system like those
appearing in nuclear physics \cite{Guhr} or in systems of interacting
fermions like quantum computers \cite{Guhr}. A well-known
method, developed by Wigner in the $1950$s, for modeling
the energy spectrum of a complex quantum system containing many
states interacting with each other, consists of assuming that
their Hamiltonian is a random matrix \cite{Matal,Porter,Mello}. Here, we
suppose that the Hamiltonian of the environment is a Gaussian orthogonal
random matrix (GORM). The interaction between the spin and the environment
is given by a coupling operator which is the product
of a system and a environment operators. The latter
is also represented by a GORM because of
its complex interaction with the many degrees of
freedom of the environment.
Such random-matrix models have recently turned out to be of great relevance
for the discussion of relaxation and dissipation in quantum systems
\cite{Cohen,Pereyra1,Pereyra2,Lutz,Lebowitz}.

The spin-GORM model can therefore be considered as
a particular case of the spin-environment model in which the environment
operators
are GORM. The Hamiltonian of the spin-GORM model is thus given by
\begin{equation}
\hat{H}_{\rm tot} = \frac{\Delta}{2} \hat{\sigma}_{z} + \hat{H}_{B} +
\lambda \hat{\sigma}_x \hat{B}
\label{hamiltspingoe}
\end{equation}
where the Hamiltonian of the environment is
\begin{equation}
\hat{H}_{B}=\frac{1}{\sqrt{8N}}\; \hat{X} \; ,
\end{equation}
and the environment coupling operator by
\begin{equation}
\hat{B}=\frac{1}{\sqrt{8N}}\; \hat{X}' \; .
\end{equation}
$\hat{X}$ and $\hat{X}'$ are two statistically independent
$\frac{N}{2} \times \frac{N}{2}$ random matrices belonging
to the Gaussian orthogonal ensemble (GOE) of probability density
\begin{equation}
p(\hat X) = {\cal C} \; \exp\left( -a_{\hat{X}} {\rm Tr} \hat X^2 \right) \; ,
\end{equation}
with $a_{\hat{X}}=\frac{1}{2}$ and a normalization constant ${\cal C}$.
$N/2$ is the
number of states of the environment. The off-diagonal and diagonal
elements of $\hat{X}$ are independent Gaussian random numbers with
mean zero and standard deviations
$\sigma_{\rm off-diag}=1$ and
$\sigma_{\rm diag}=\sqrt{2}$ respectively.

In the limit $N \to \infty$, the density of states of the environment gets
smooth
and can be calculated by an average over the random-matrix ensemble
\begin{equation}
n(\epsilon)=\overline{\sum_{b=1}^{N/2} \delta(\epsilon-E_b)}\; .
\end{equation}
It is known that the GOE level density is given by the Wigner semicircle law:
\begin{equation}
n(\epsilon) =
\left\{
\begin{array}{ll}
\frac{4N}{\pi} \sqrt{\frac{1}{4}-\epsilon^2} & {\rm if}
\quad \vert \epsilon \vert < \frac{1}{2}  \\ 0 & {\rm if}
\quad \vert \epsilon \vert \geq \frac{1}{2} \; .
\end{array}
\right.
\label{semi-circ}
\end{equation}
The random matrices are normalized so that
the level density of the environment has a width equal to unity.
To simplify the notations in the following, we use the convention
\begin{equation}
\sqrt{x} \equiv \left\{
\begin{array}{ll}
\sqrt{x} & {\rm if} \quad 0 < x \\
0        & {\rm if} \quad x \leq 0 \; .
\end{array}
\right.
\label{sqrt}
\end{equation}

For the following, we also need to evaluate the function
$F(\epsilon,\epsilon')$
for our random-matrix model. For this purpose, we need the random-matrix
average
of the quantity (\ref{fctcorrmicpauli}).
Since, in the GOE, we have that
\begin{eqnarray}
\overline{\vert \langle b \vert \hat B\vert b\rangle \vert^2} &=&
\frac{1}{4N} \; , \\
\overline{\vert \langle b \vert \hat B\vert b'\rangle \vert^2} &=&
\frac{1}{8N} \; ,
\end{eqnarray}
we find that
\begin{eqnarray}
& & \overline{{\rm Tr}_{B} \delta(\epsilon-\hat{H}_B) \hat{B}
\delta(\epsilon'-\hat{H}_B) \hat{B} } \nonumber \\
&=& \overline{\sum_{b,b'} \delta(\epsilon-E_b) \delta(\epsilon'-E_{b'})
\vert \langle b \vert \hat B\vert b'\rangle \vert^2} \nonumber \\
&=& \sum_{b} \overline{\delta(\epsilon-E_b) \delta(\epsilon'-E_{b})}
\overline{\vert \langle b \vert \hat B\vert b\rangle \vert^2}
+ \sum_{b\ne b'} \overline{\delta(\epsilon-E_b) \delta(\epsilon'-E_{b'})}
\overline{\vert \langle b \vert \hat B\vert b'\rangle \vert^2} \nonumber \\
&=& \frac{1}{4N} \overline{\sum_{b} \delta(\epsilon-E_b)
\delta(\epsilon'-E_{b})}
+ \frac{1}{8N} \overline{\sum_{b\ne b'} \delta(\epsilon-E_b)
\delta(\epsilon'-E_{b'})} \nonumber \\
&=& \frac{1}{8N} \overline{\sum_{b} \delta(\epsilon-E_b)
\delta(\epsilon'-E_{b})}
+ \frac{1}{8N} \overline{\sum_{b, b'} \delta(\epsilon-E_b)
\delta(\epsilon'-E_{b'})} \nonumber \\
&\simeq& \frac{1}{8N} \delta(\epsilon-\epsilon') n(\epsilon) + \frac{1}{8N}
n(\epsilon) n(\epsilon') \; .
\end{eqnarray}
In the limit $N\to\infty$, the first term becomes negligible in front of
the second term
so that the comparison with Eq. (\ref{fctcorrmicpauli}) shows that for the
spin-GORM model,
\begin{equation}
F(\epsilon,\epsilon') \simeq \frac{1}{8N} \; .
\label{F-GORM}
\end{equation}

\begin{figure}[P]
\centering
\rotatebox{0}{\scalebox{0.6}{\includegraphics{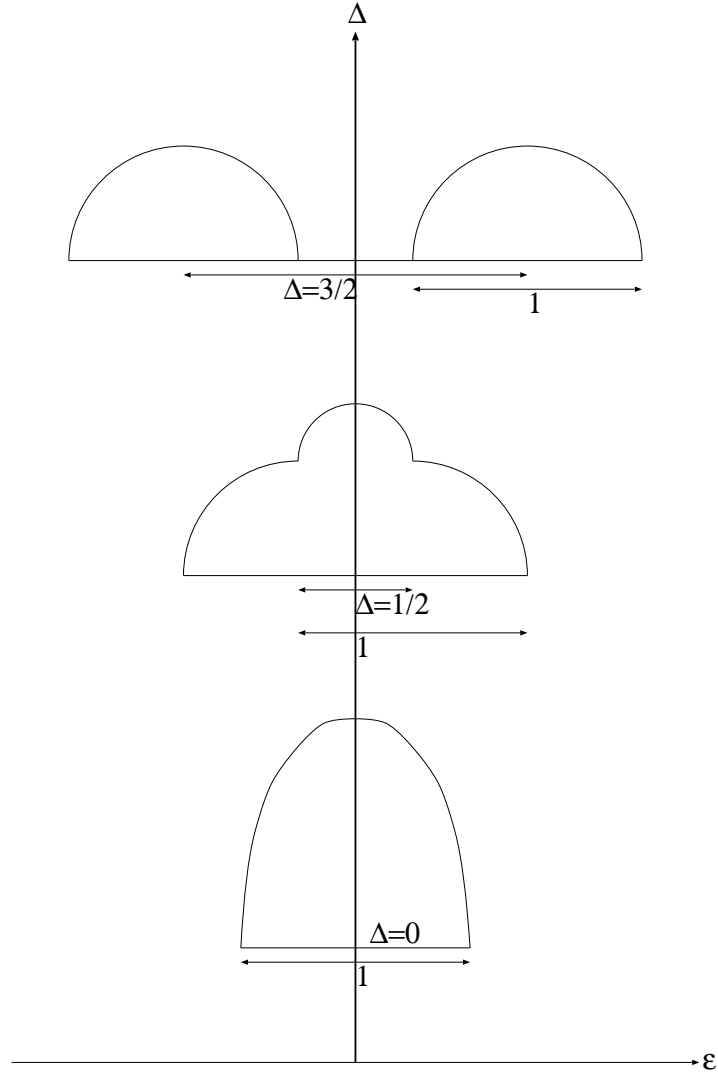}}} \\
\caption{Schematic representation of the smooth density of states (DOS)
of the unperturbed spin-GORM model ($\lambda=0$) for different values
of the energy splitting $\Delta$ of the spin. The horizontal axis
is the environment energy $\epsilon$ while the
vertical axis is the energy splitting $\Delta$.  In the lower
and central parts, the splitting $\Delta$ is smaller than the width of the
environment DOS.
In the upper part, the splitting $\Delta$ is larger than the width of the DOS.
\label{shemdensdel}}
\end{figure}

The total system contains $N$ states.
The unperturbed density of states of the total system is
schematically depicted in Fig. \ref{shemdensdel}, for $\lambda=0$.
The model has different regimes whether the splitting $\Delta$
between the two levels of the spin is larger or smaller
than the width of the environment level density.
The spin-GORM model can describe a large variety of physical situations.
In the present paper we focus on the perturbative regimes ($\lambda \ll
\hat{H}_B$). When
$\Delta$ is larger than the width of the semicircular
density of states of the environment, we are in a highly
non-Markovian regime. The dynamics of the system is faster than that of
the environment. On the other hand, when $\Delta$ is smaller than unity,
we are in a Markovian regime because the dynamics of the
environment is much faster then the one of the system.

Now, we apply our master equation and the Redfield equation to
the spin-GORM model in both their Markovian and
non-Markovian versions.

%%%%%%%%%%%%%%%%%%%%%%%%%%%%%%%%%%%%%%%%%%%%%%%%%%%%%%%%%%%%%%%%%%%
\subsection{Using our master equation}

We now apply our master equation (\ref{our.nonMarkov.master.eq}) to the
spin-GORM model.
Using Eqs. (\ref{paulispinbosNM11})-(\ref{paulispinbosNM1-1}),
(\ref{semi-circ}), and (\ref{F-GORM}), we get
the non-Markovian equations
\begin{eqnarray}
\dot{P}_{++}(\epsilon;t) &=& - \frac{\lambda^2}{\pi} P_{++}(\epsilon;t)
\int d\epsilon'
\sqrt{\frac{1}{4}-{\epsilon'}^2} \;
\frac{\sin(\epsilon-\epsilon'+\Delta)t}{(\epsilon-\epsilon'+\Delta)}
\nonumber \\
& & + \frac{\lambda^2}{\pi} \sqrt{\frac{1}{4}-\epsilon^2} \int d\epsilon'
P_{--}(\epsilon';t) \;
\frac{\sin(\epsilon-\epsilon'+\Delta)t}{(\epsilon-\epsilon'+\Delta)} \;
,\nonumber \\
\label{paulispingoeNM11}
\end{eqnarray}
\begin{eqnarray}
\dot{P}_{--}(\epsilon;t) &=& - \frac{\lambda^2}{\pi} P_{--}(\epsilon;t)
\int d\epsilon'
\sqrt{\frac{1}{4}-{\epsilon'}^2} \;
\frac{\sin(\epsilon-\epsilon'-\Delta)t}{(\epsilon-\epsilon'-\Delta)}
\nonumber \\
& & + \frac{\lambda^2}{\pi} \sqrt{\frac{1}{4}-\epsilon^2} \int d\epsilon'
P_{++}(\epsilon';t) \;
\frac{\sin(\epsilon-\epsilon'-\Delta)t}{(\epsilon-\epsilon'-\Delta)} \;
,\nonumber \\
\label{paulispingoeNM-1-1}
\end{eqnarray}
\begin{eqnarray}
\dot{P}_{+-}(\epsilon;t) &=& -i \Delta \; P_{+-}(\epsilon;t) \nonumber \\
& & - \frac{\lambda^2}{2\pi} P_{+-}(\epsilon;t) \int d\epsilon'
\sqrt{\frac{1}{4}-{\epsilon'}^2} \int_{0}^{t} d\tau
\left[e^{i (\epsilon-\epsilon'+\Delta) \tau}+e^{-i
(\epsilon-\epsilon'-\Delta) \tau}\right] \nonumber \\
& & + \frac{\lambda^2}{2\pi} \sqrt{\frac{1}{4}-\epsilon^2} \int d\epsilon'
P_{-+}(\epsilon';t) \int_{0}^{t} d\tau
\left[ e^{i (\epsilon-\epsilon'-\Delta) \tau}+e^{-i
(\epsilon-\epsilon'+\Delta) \tau}\right] \; . \nonumber \\
\label{paulispingoeNM1-1}
\end{eqnarray}
We notice that the equations for the populations are decoupled from the
ones for
the quantum coherences.

Performing the \textit{Markovian approximation} and using
$\lim_{t\to\infty}\frac{\sin\omega
t}{\omega}=\pi\delta(\omega)$ and Eq. (\ref{int0inf}), we get
\begin{equation}
\dot{P}_{++}(\epsilon;t) = - \lambda^2 \sqrt{\frac{1}{4}-(\epsilon+\Delta)^2}
\; P_{++}(\epsilon;t)
+\lambda^2 \sqrt{\frac{1}{4}-\epsilon^2}
\; P_{--}(\epsilon+\Delta;t) \; , \label{pauli11}
\end{equation}
\begin{equation}
\dot{P}_{--}(\epsilon;t) = - \lambda^2 \sqrt{\frac{1}{4}-(\epsilon-\Delta)^2}
\; P_{--}(\epsilon;t)
+\lambda^2 \sqrt{\frac{1}{4}-\epsilon^2}
 \; P_{++}(\epsilon-\Delta;t) \; ,\label{pauli-1-1}
\end{equation}
\begin{eqnarray}
\dot{P}_{+-}(\epsilon;t) &=&  -i \Delta \; P_{+-}(\epsilon;t) \nonumber \\
& & +i \; \frac{\lambda^2}{\pi} \int d\epsilon' \;
{\cal P}\frac{\Delta}{(\epsilon-\epsilon')^2-\Delta^2}
\left[\sqrt{\frac{1}{4}-{\epsilon'}^2}
\; P_{+-}(\epsilon;t) + \sqrt{\frac{1}{4}-\epsilon^2}
\; P_{-+}(\epsilon';t) \right] \nonumber \\
& & - \frac{\lambda^2}{2} \left[ \sqrt{\frac{1}{4}-(\epsilon+\Delta)^2}+
\sqrt{\frac{1}{4}-(\epsilon-\Delta)^2}\right] P_{+-}(\epsilon;t) \nonumber \\
& & + \frac{\lambda^2}{2} \sqrt{\frac{1}{4}-\epsilon^2}
\left[ P_{-+}(\epsilon+\Delta;t)
+ P_{-+}(\epsilon-\Delta;t) \right] \; ,\nonumber \\
\label{pauli1-1}
\end{eqnarray}
where the expressions and integrals over energy extend over the interval of
definition
of the level density $n(E)$ and of the distributions $P_{ss'}(E;t)$
which is always $-1/2< E < +1/2$, $E$ being the argument of these functions.

We now focus our attention on the evolution of the populations.

\begin{figure}[P]
\centering
\rotatebox{0}{\scalebox{0.8}{\includegraphics{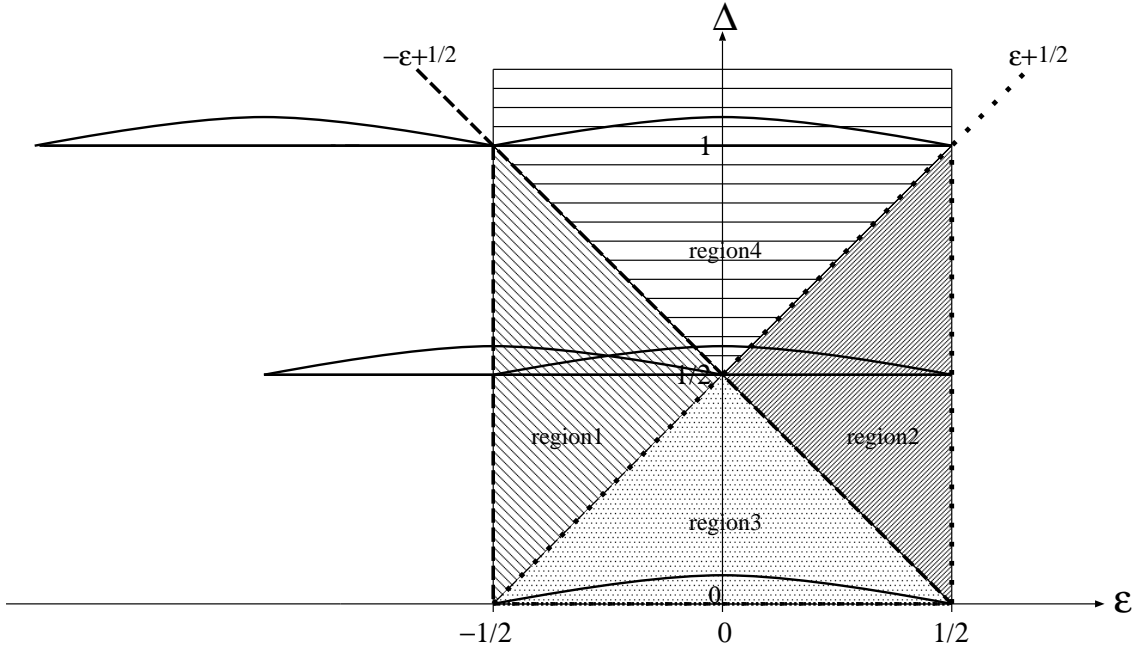}}} \\
\caption{Schematic representation of the different regimes
of the spin-GORM model for situations where the
initial state of the spin is $s=+1$, in the plane of
the environment energy $\epsilon$ versus the spin energy splitting $\Delta$.
The different regions correspond to different values
of the functions $n(\epsilon)$,
$n(\epsilon+\Delta)$, and
$n(\epsilon-\Delta)$, where $n(E)$ denotes the DOS defined by the
semicircle law (\ref{semi-circ}). One can take a value of the environment
energy
anywhere between $\epsilon=-\frac{1}{2}$ and
$\epsilon=\frac{1}{2}$, where $n(\epsilon) \ne 0$.
In region $1$: $n(\epsilon+\Delta)\ne 0$ and
$n(\epsilon-\Delta)=0$. In region $2$:
$n(\epsilon+\Delta)=0$ and
$n(\epsilon-\Delta)\ne 0$. In region $3$:
$n(\epsilon+\Delta)\ne 0$ and
$n(\epsilon-\Delta)\ne 0$. In region $4$:
$n(\epsilon+\Delta)=0$ and
$n(\epsilon-\Delta)=0$. This figure is useful
to evaluate equations such as Eqs.
(\ref{paulispingoeMZinfini}) and (\ref{paulispingoeMrate}) and
(\ref{redspingoeMZinfini}) and (\ref{redspingoeMrate}).
\label{schematriangle}}
\end{figure}

We see from Eqs. (\ref{pauli11}) and (\ref{pauli-1-1}) that the transitions
conserve
the total energy of the system and environment so that the transitions
occur between the only two energies $\epsilon$ and $\epsilon+\Delta$ of the
environment.
As a consequence, the quantity
\begin{equation}
P(\epsilon;t) \equiv P_{++}(\epsilon;t) + P_{--}(\epsilon+\Delta;t) =
P(\epsilon;0) 
\end{equation}
is a constant of the motion for each energy $\epsilon$ of the environment,
as already noticed
with Eq. (\ref{cst}).  The difference (\ref{def.Z}) of populations obeys
the differential equation (\ref{eq.Z}). If the initial distributions
$P(\epsilon';0)$ and $Z(\epsilon';0)$ are
Dirac delta distributions centered on the initial energy $\epsilon$:
\begin{eqnarray}
P(\epsilon';0) &=& \delta(\epsilon'-\epsilon) \; , \\
Z(\epsilon';0) &=& \delta(\epsilon'-\epsilon) \; z_{\rm Pauli}(0) \; .
\end{eqnarray}
The $z$ component of the spin defined as
\begin{equation}
z_{\rm Pauli}(t) = \int d\epsilon \; Z(\epsilon;t) 
\end{equation}
obeys the same differential equation as the distribution $Z(\epsilon;t)$,
\begin{eqnarray}
\dot z_{\rm Pauli}(t) &=& \lambda^2 \left[ \sqrt{\frac{1}{4}-\epsilon^2}-
\sqrt{\frac{1}{4}-(\epsilon+\Delta)^2}
\right] \nonumber \\
& & - \lambda^2 \left[ \sqrt{\frac{1}{4}-\epsilon^2} +
\sqrt{\frac{1}{4}-(\epsilon+\Delta)^2}
\right] z_{\rm Pauli}(t) \; .
\label{z_Pauli.eq}
\end{eqnarray}
The solution of Eq. (\ref{z_Pauli.eq}) is given by
\begin{eqnarray}
z_{\rm Pauli}(t)= z_{\rm Pauli}(\infty) + \left[ z_{\rm Pauli}(0)- z_{\rm
Pauli}(\infty) \right] e^{-
\gamma_{\rm Pauli} t} \; ,
\label{paulispingoeMZt}
\end{eqnarray}
with the asymptotic equilibrium value
\begin{equation}
z_{\rm
Pauli}(\infty)=\frac{\sqrt{\frac{1}{4}-\epsilon^2}-\sqrt{\frac{1}{4}-
(\epsilon+\Delta)^2}}
{\sqrt{\frac{1}{4}-\epsilon^2} +\sqrt{\frac{1}{4}-(\epsilon+\Delta)^2}} \; ,
\label{paulispingoeMZinfini}
\end{equation}
and the relaxation rate
\begin{equation}
\gamma_{\rm Pauli}= \lambda^2 \left[ \sqrt{\frac{1}{4}-\epsilon^2}+
\sqrt{\frac{1}{4}-(\epsilon+\Delta)^2} \right] \; . \label{paulispingoeMrate}
\end{equation}
With the convention (\ref{sqrt}), the expressions are nonvanishing only
over the interval
of definition of their argument.
Figure \ref{schematriangle} helps us to represent the different values that
take Eqs. (\ref{paulispingoeMZinfini})
and (\ref{paulispingoeMrate}) in the space of the environment energy
$\epsilon$ and of the splitting energy
$\Delta$ of the two-level system.

%%%%%%%%%%%%%%%%%%%%%%%%%%%%%%%%%%%%%%%%%%%%%%%%%%%%%%%%%%%%%%%%%%%
\subsection{Using the Redfield equation }

For the spin-GORM model, the correlation function (\ref{defcorrel}) can be
calculated by performing a GOE average.
Using the level density (\ref{semi-circ}) and the value (\ref{F-GORM}), the
microcanonical correlation function
(\ref{fctcorrmicred}) becomes
\begin{eqnarray}
\alpha(\tau,\epsilon)
&=& \int d\epsilon' \; n(\epsilon') \; F(\epsilon,\epsilon') \;
e^{i(\epsilon-\epsilon')\tau} \nonumber \\
&\simeq& \int_{-\frac{1}{2}}^{+\frac{1}{2}} d\epsilon' \; \frac{4N}{\pi}
\sqrt{\frac{1}{4}-{\epsilon'}^2} \; \frac{1}{8N} \;
e^{i(\epsilon-\epsilon')\tau} \nonumber \\
&=& \frac{J_1(\frac{\tau}{2})}{4\tau} \; e^{i \epsilon \tau} \; ,
\label{fctcorrmicredGOE}
\end{eqnarray}
in the limit $N\to\infty$, where $J_1(t)$ is the Bessel function of the
first kind.

Therefore, using the Redfield equation
(\ref{redspinboscorrNM11})-(\ref{redspinboscorrNM1-1})
and the microcanonical correlation function of the spin-GORM
model, we get
\begin{eqnarray}
\dot\eta_{++}(t)&=&
-\lambda^2 \; \eta_{++}(t)
\int_{0}^{t} d\tau \cos[(\epsilon+\Delta)\tau]
\frac{J_1(\frac{\tau}{2})}{2\tau} \nonumber \\
& &+\lambda^2 \; \eta_{--}(t)
\int_{0}^{t} d\tau \cos[(\epsilon-\Delta)\tau]
\frac{J_1(\frac{\tau}{2})}{2\tau} \; ,
\label{redspingoeNM11}
\end{eqnarray}
\begin{eqnarray}
\dot\eta_{--}(t)&=&
-\lambda^2 \; \eta_{--}(t)
\int_{0}^{t} d\tau \cos[(\epsilon-\Delta)\tau]
\frac{J_1(\frac{\tau}{2})}{2\tau} \nonumber \\
& &+\lambda^2 \; \eta_{++}(t)
\int_{0}^{t} d\tau \cos[(\epsilon+\Delta)\tau]
\frac{J_1(\frac{\tau}{2})}{2\tau} \; ,
\label{redspingoeNM-1-1}
\end{eqnarray}
\begin{eqnarray}
\dot\eta_{+-}(t)&=& -i \Delta \; \eta_{+-}(t) \nonumber \\
& & -\lambda^2 \; \eta_{+-}(t)
\int_{0}^{t} d\tau e^{i\Delta\tau} \cos(\epsilon \tau)
\frac{J_1(\frac{\tau}{2})}{2\tau} \nonumber \\
& & +\lambda^2 \; \eta_{-+}(t)
\int_{0}^{t} d\tau e^{-i\Delta\tau} \cos(\epsilon \tau)
\frac{J_1(\frac{\tau}{2})}{2\tau} \; .
\label{redspingoeNM1-1}
\end{eqnarray}

These are the non-Markovian Redfield equations for the spin-GORM model.

The Markovian Redfield equations for the spin-GORM model take the following
forms:
\begin{eqnarray}
\dot\eta_{++}(t)&=&
-\lambda^2 \sqrt{\frac{1}{4}-(\epsilon+\Delta)^2}
\;
\eta_{++}(t) \nonumber \\
& &+\lambda^2 \sqrt{\frac{1}{4}-(\epsilon-\Delta)^2}
\;
\eta_{--}(t) \; ,
\label{redspingoeM11}
\end{eqnarray}
\begin{eqnarray}
\dot\eta_{--}(t)&=&
-\lambda^2 \sqrt{\frac{1}{4}-(\epsilon-\Delta)^2}
\;
\eta_{--}(t) \nonumber \\
& &+\lambda^2 \sqrt{\frac{1}{4}-(\epsilon+\Delta)^2}
\;
\eta_{++}(t) \; ,
\label{redspingoeM-1-1}
\end{eqnarray}
\begin{eqnarray}
\dot\eta_{+-}(t)&=& -i \Delta \; \eta_{+-}(t) \nonumber \\
& & + i \; \frac{\lambda^2}{\pi} \int d\epsilon' \;
\sqrt{\frac{1}{4}-{\epsilon'}^2}
\; {\cal P}\frac{\Delta}{(\epsilon-\epsilon')^2-\Delta^2} \nonumber \\
& & \qquad\qquad\qquad\times \;
\left[ \eta_{+-}(t) +
\eta_{-+}(t) \right] \nonumber \\
& & -\frac{\lambda^2}{2} \left[ \sqrt{\frac{1}{4}-(\epsilon+\Delta)^2}
+ \sqrt{\frac{1}{4}-(\epsilon-\Delta)^2} \right] \nonumber \\
& & \qquad\qquad\qquad \times \left[
\eta_{+-}(t)-\eta_{-+}(t) \right] \; .
\nonumber \\
\label{redspingoeM1-1}
\end{eqnarray}

We focus on the evolution of the populations.

The population of the two-level system is controlled by the $z$ component of
the spin by Eq. (\ref{redspinbosonMZtdef}). According to the Markovian
Redfield equations (\ref{redspingoeM11})
and (\ref{redspingoeM-1-1}), the time evolution of the $z$ component is
given by
Eq. (\ref{redspinbosonMZt}) with
with the equilibrium value
\begin{equation}
z_{\rm Redfield}(\infty)=\frac{\sqrt{\frac{1}{4}-(\epsilon-\Delta)^2}
-\sqrt{\frac{1}{4}-(\epsilon+\Delta)^2}}
{\sqrt{\frac{1}{4}-(\epsilon-\Delta)^2}
+\sqrt{\frac{1}{4}-(\epsilon+\Delta)^2}} \; ,
\label{redspingoeMZinfini}
\end{equation}
and the relaxation rate
\begin{equation}
\gamma_{\rm Redfield}=\lambda^2 \left[ \sqrt{\frac{1}{4}-(\epsilon-\Delta)^2}
+\sqrt{\frac{1}{4}-(\epsilon+\Delta)^2}\right] \; .
\label{redspingoeMrate}
\end{equation}
Figure \ref{schematriangle} depicts the different regimes predicted by this
equation in the space
of the environment energy $\epsilon$ and the splitting energy $\Delta$.

We observe that in the limit $\Delta\to 0$, where the energy scale of the
system
is much smaller than the one of the environment, both our master equation and
the Redfield equation predict a similar value for the relaxation rate.
However, differences appear for the value of the
asymptotic value of the $z$ component of the spin.
As we explained here above, the reason is that our master equation
is consistent with energy conservation in the total system albeit
the Redfield equation is not.  This problem spoils the applicability
of the Redfield master equation if the environment is not arbitrarily
large as we shall see in the next section.

%%%%%%%%%%%%%%%%%%%%%%%%%%%%%%%%%%%%%%%%%%%%%%%%%%%%%%%%%%%%%%%%%%%%%%%%%%%%%%%%
%
%%%%%%%%%%%%%%%%%%%%%%%%%%%%%%%%%%%%%%%%%%%%%%%%%%%%%%%%%%%%%%%%%%%%%%%%%%%%%%%%
%
\section{Numerical results and discussion}
\label{numres}

The purpose of the present section is to compare the different master equations
with exact numerical results obtained for
the relaxation of the $z$ component of the spin due to the
interaction with its environment in the spin-GORM model. The
initial condition of the spin is always the state $\vert + \rangle$. The
environment is always taken in a microcanonical distribution at a
given energy $\epsilon$. The width of the energy shell of
this microcanonical distribution is always equal to
$\delta \epsilon = 0.05$.

A general comment is here in order concerning the applicability of
a master equation to a quantum system with a discrete
energy spectrum.  Indeed, beyond a time longer than the Heisenberg time
(which is defined as the level density of the total system),
quantum beats and recurrences appear due to the discreteness of
the energy spectrum.  Only, the decay before the Heisenberg time
can be compared with the prediction of a quantum master equation.
It turns out that the further condition $N\lambda^2 >1$ should also be
satisfied,
which requires that the coupling parameter should not be too small with
respect to the mean level
spacing which goes as $1/N$.  If this condition is not satisfied
($N\lambda^2 <1$)
the time evolution of individual systems present large quantum oscillations
which widely deviate from the prediction of the master equation.
On the other hand, if $N\lambda^2 >1$, the deviations with respect to
the predictions of the master equation are smaller
than the signal itself and tend to decrease as $N\to\infty$ \cite{EspoGasp}.
In the limit $N\to\infty$, the decay curve which is the solution
of the master equation is approximately followed by a majority of realizations
of the process by individual systems.  In the figures given here,
these deviations are not seen because of an averaging of the signal over
$\chi=10$
individual systems.  Besides the condition $N\lambda^2 >1$,
the coupling parameter should also be small enough to justify
the perturbative treatment, typically $\lambda < 0.3$.

The equations we are comparing are the following:

$\bullet$ The von Neumann equation describes the $z$ component of the spin
using
Eq. (\ref{vonNeumann}) with the Hamiltonian
(\ref{hamiltspingoe}).  Averaging is carried out with $\chi$ realizations
of the
GORM Hamiltonian.  This calculation does not involve any approximation
and, therefore, gives the exact solution of the problem. All the
following equations will be compared to this one.

$\bullet$ The most general non-Markovian version of our master equation
(\ref{our.nonMarkov.master.eq}) using Eqs. (\ref{paulispingoeNM11}) and
(\ref{paulispingoeNM-1-1}), which we refer to as the Pauli non-Markovian
(NM) equation.

$\bullet$ The Markovian version (\ref{ourMarkovmastereq}) of our master
equation,
which we refer to as the Pauli Markovian (M) equation. For the spin-GORM model,
this equation is given by Eqs. (\ref{pauli11}) and (\ref{pauli-1-1}) and
its solutions
by Eqs. (\ref{paulispingoeMZt})-(\ref{paulispingoeMrate}).

$\bullet$ The Redfield non-Markovian equation
(\ref{Redfield.nonMarkov.master.eq}) is given by
Eqs. (\ref{redspingoeNM11}) and (\ref{redspingoeNM-1-1}) for the spin-GORM
model.

$\bullet$ The standard Redfield Markovian equation
(\ref{Redfield.Markov.master.eq}) is given
by Eqs. (\ref{redspingoeM11}) and (\ref{redspingoeM-1-1})
and its solutions by Eqs. (\ref{redspinbosonMZt}) with
Eqs. (\ref{redspingoeMZinfini}) and (\ref{redspingoeMrate}).

\begin{figure}[P]
\centering
\rotatebox{0}{\scalebox{0.6}{\includegraphics{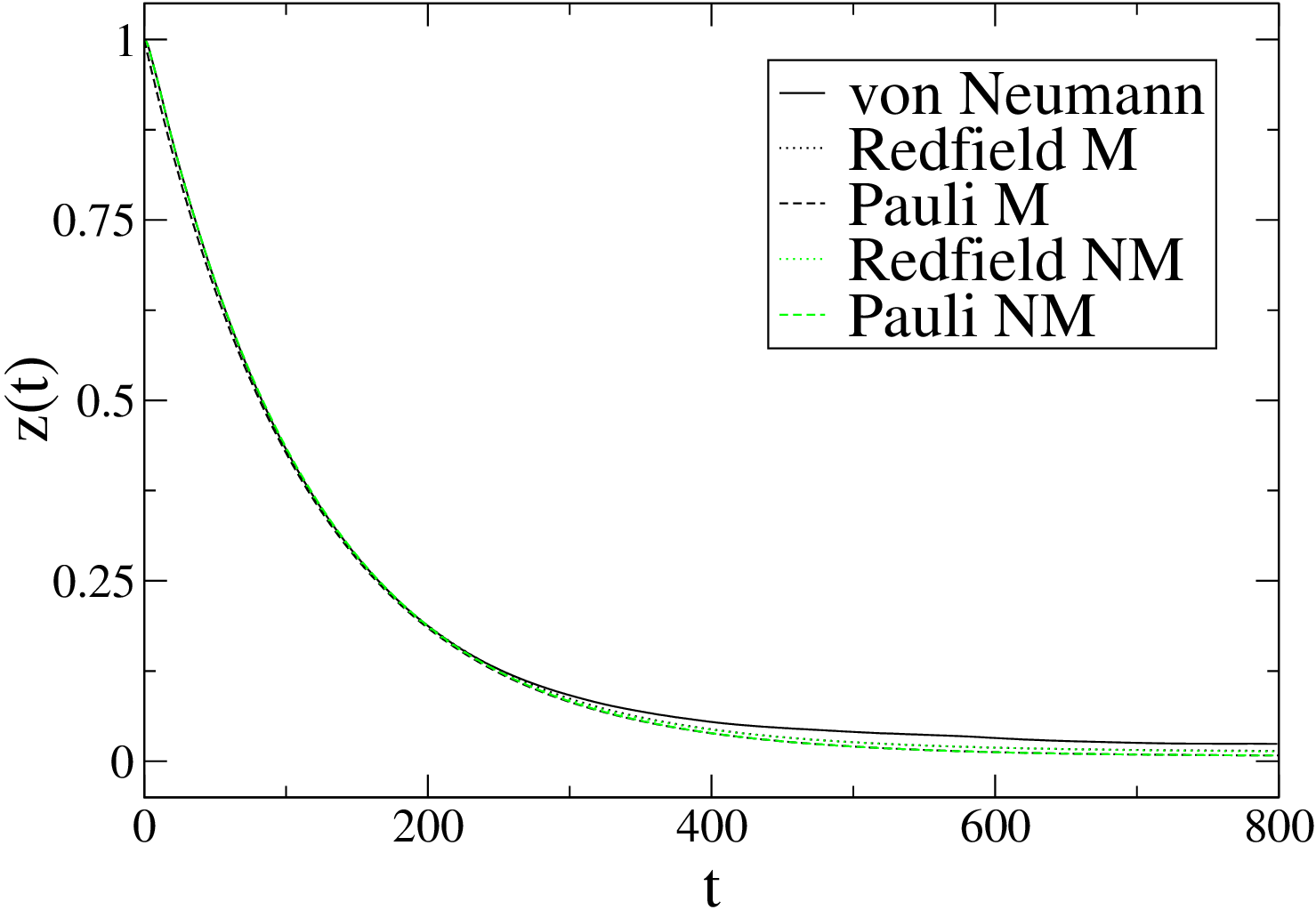}}} \\
\caption{Relaxation of the $z$ component of the spin for the
spin-GORM model for a very small spin energy splitting $\Delta=0.01$ with
$\lambda=0.1$,
$\epsilon=0.25$, $N=2000$, and $\chi=10$. The exact curve is given by
integrating
the von Neumann equation, which is compared with the solutions of the
Pauli and Redfield Markovian (M) and non-Markovian (NM) equations. We see
that all the
perturbative equations give similar results in the present case.
\label{z-d=0.01}}
\end{figure}

\begin{figure}[P]
\centering
\rotatebox{0}{\scalebox{0.6}{\includegraphics{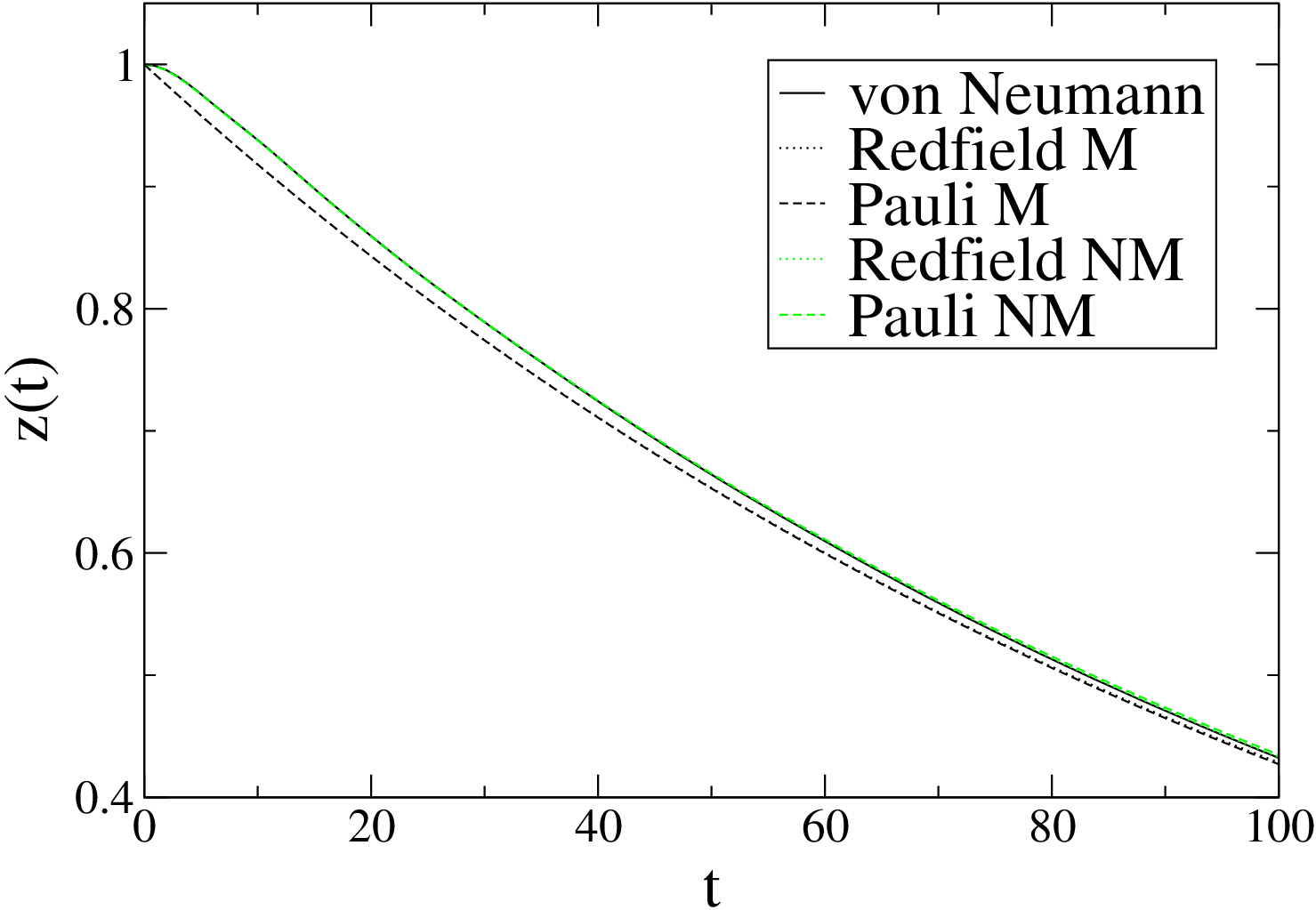}}}\\
\caption{Relaxation of the $z$ component of the spin for the
spin-GORM model in the same conditions $\Delta=0.01$, $\lambda=0.1$,
$\epsilon=0.25$, $N=2000$, and $\chi=10$ as in Fig. \ref{z-d=0.01}
in order to show that, on a short time scale of the order of
the correlation time of the environment $\tau_{\rm corr} \simeq 10$,
the non-Markovian equations (denoted by NM) describe very
accurately the dynamics although the Markovian ones (denoted by M)
is exponential and deviate from the exact behavior.
On a longer time scale (much longer than
$\tau_{\rm corr}$), the solutions of the Markovian equations join
those of the non-Markovian equations. \label{bisz-d=0.01}}
\end{figure}

\begin{figure}[P]
\centering
\rotatebox{0}{\scalebox{0.6}{\includegraphics{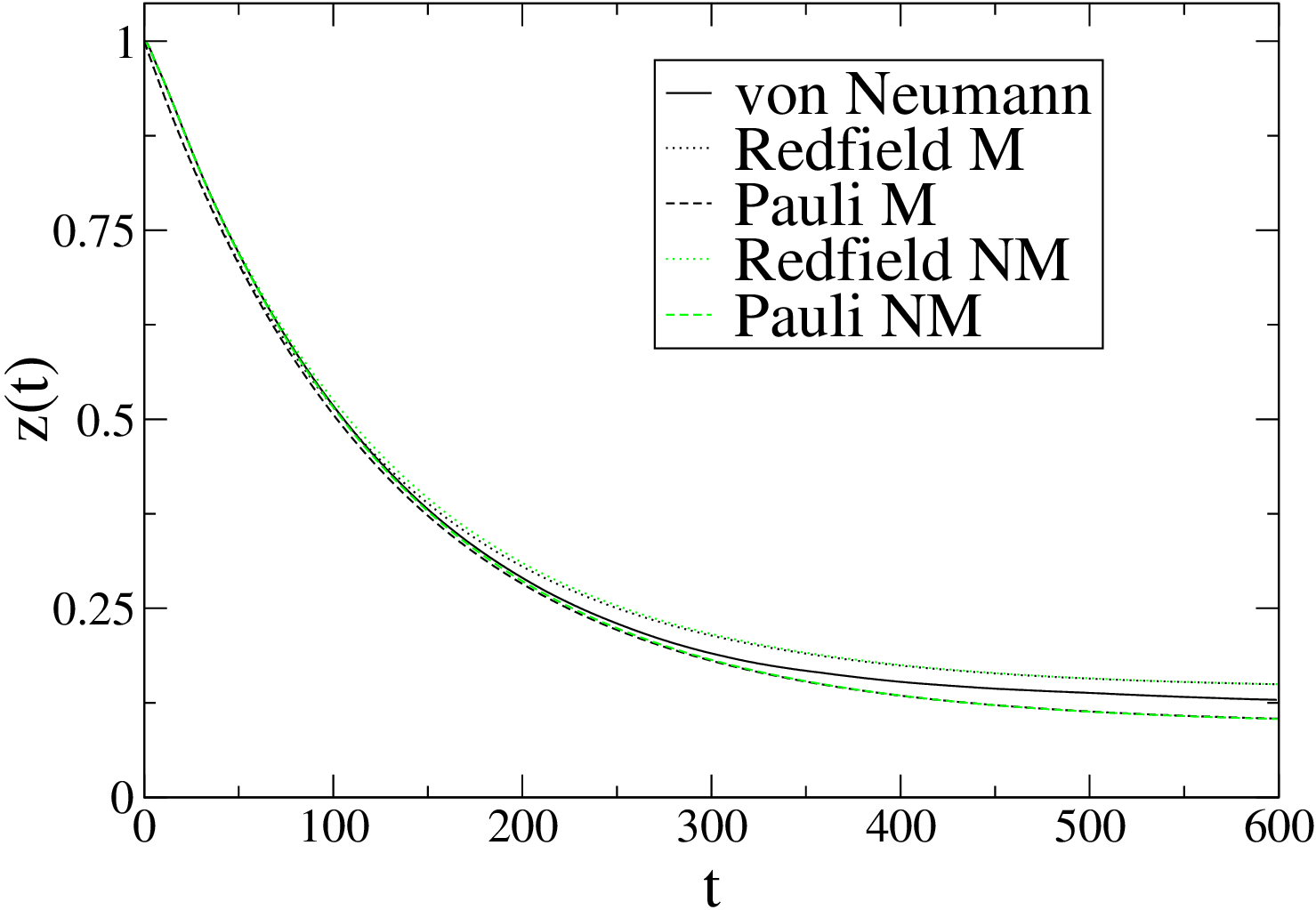}}} \\
\caption{Relaxation of the $z$ component of the spin for the
spin-GORM model for a small spin energy splitting $\Delta=0.1$ with
$\lambda=0.1$,
$\epsilon=0.25$, $N=2000$, and $\chi=10$. The
exact solution of the von Neumann equation is compared with the solutions
of the Pauli and Redfield Markovian (M) and non-Markovian (NM) equations.
We see that our master equation (Pauli NM)
gives the best results and that the solutions of the Markovian equations
remain very close to those of the non-Markovian equations. \label{z-d=0.1}}
\end{figure}

\begin{figure}[P]
\centering
\rotatebox{0}{\scalebox{0.6}{\includegraphics{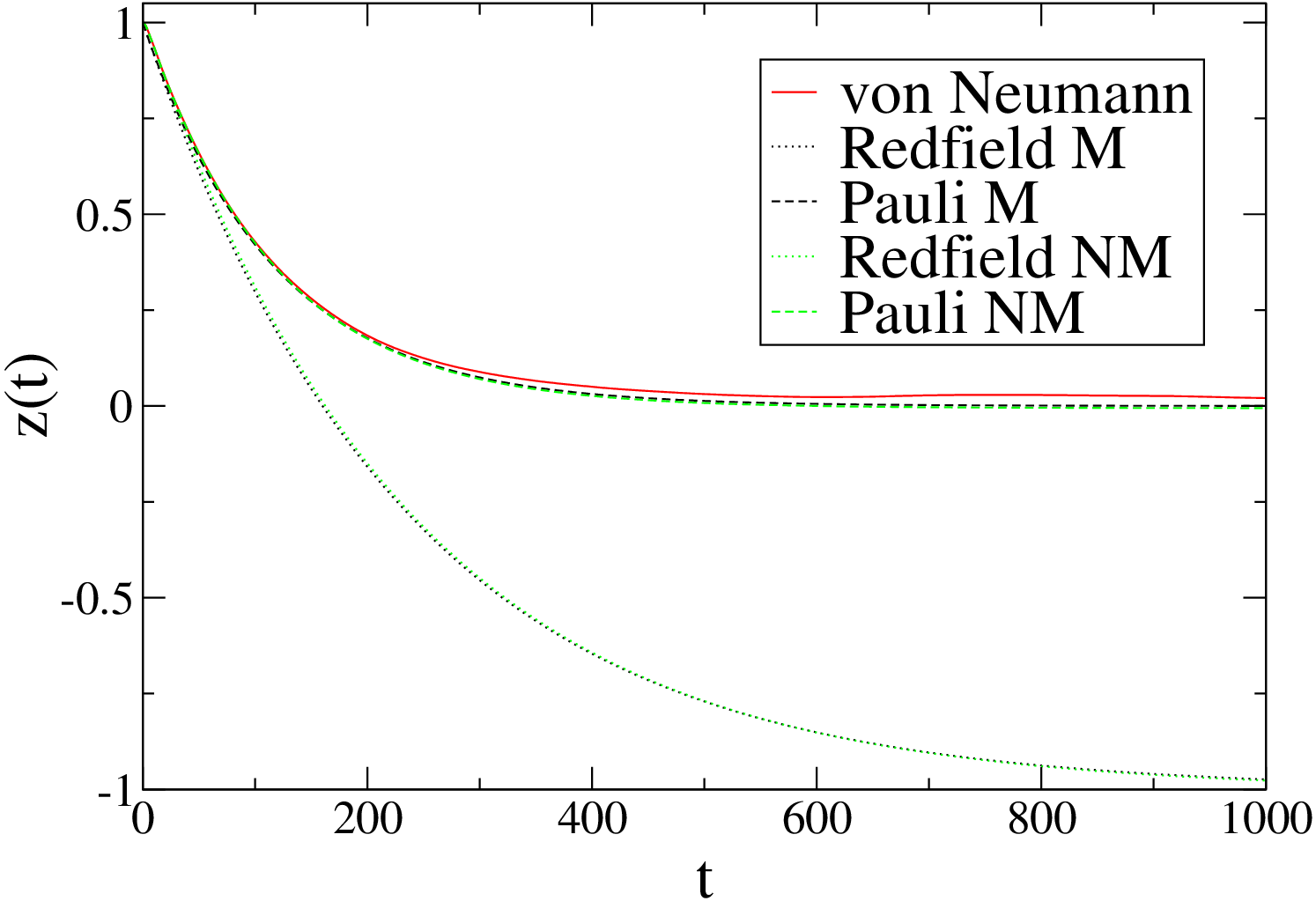}}} \\
\caption{Relaxation of the $z$ component of the spin for the
spin-GORM model for an intermediate spin energy splitting $\Delta=0.5$ with
$\lambda=0.1$,
$\epsilon=-0.25$, $N=2000$, and $\chi=10$. The
exact solution of the von Neumann equation is compared with the solutions
of the Pauli and Redfield Markovian (M) and non-Markovian (NM) equations.
Here, we see that the Redfield equations
give completely wrong results after a short time. The Pauli
equations give much better results than the Redfield ones.
\label{d=0.5l=0.1}}
\end{figure}

\begin{figure}[P]
\centering
\rotatebox{0}{\scalebox{0.6}{\includegraphics{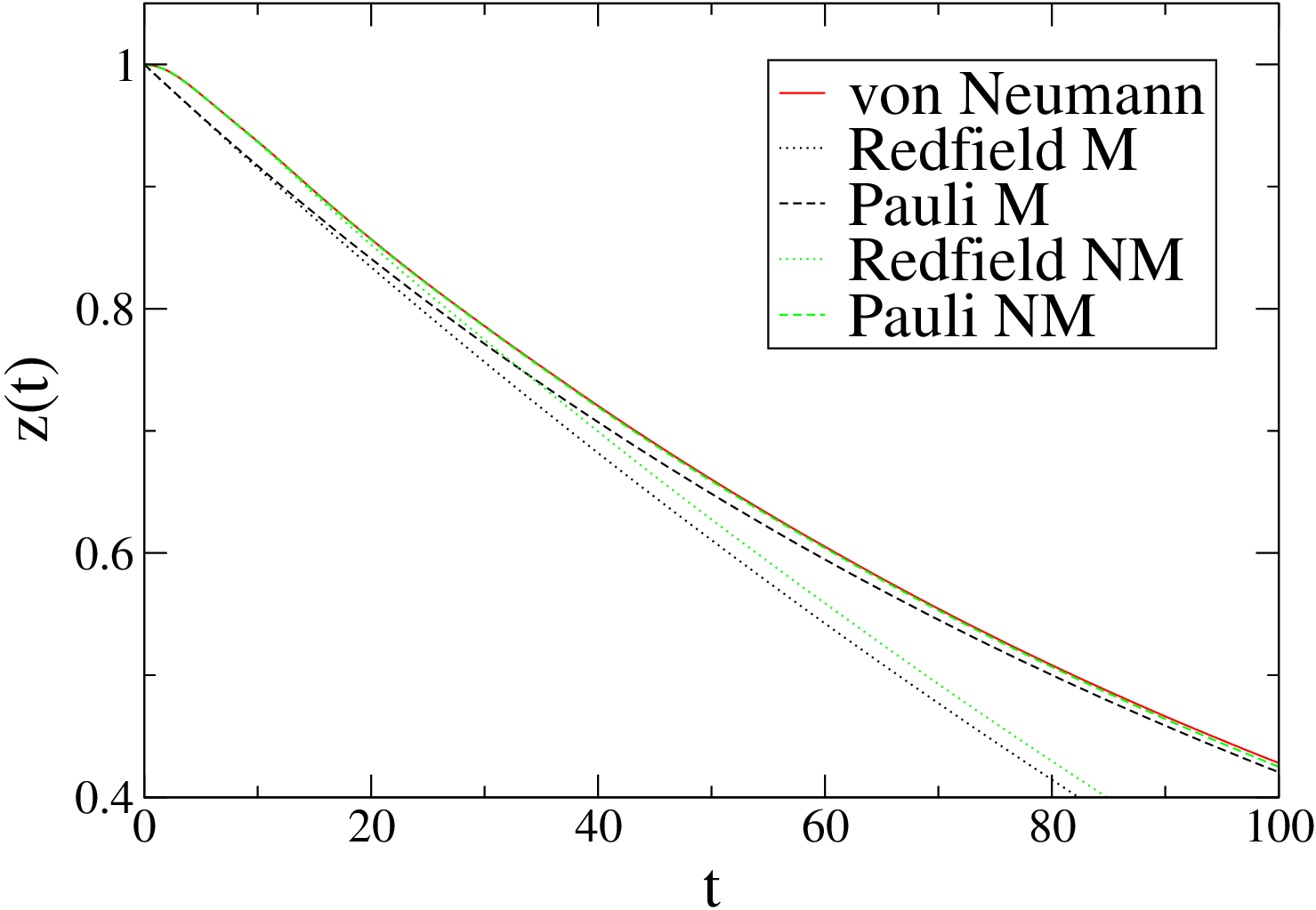}}}\\
\caption{Relaxation of the $z$ component of the spin for the
spin-GORM model in the same conditions $\Delta=0.5$, $\lambda=0.1$,
$\epsilon=-0.25$, $N=2000$, and $\chi=10$ as in Fig. \ref{d=0.5l=0.1}.
We focus here on the short time dynamics
in order to see that only the non-Markovian equations (NM)
reproduce the initial behavior of the system which is not the case
for the Markovian equations (M). After $\tau_{\rm corr}\simeq 10$, the
non-Markovian Redfield equation (Redfield NM) becomes wrong but
our master equation (Pauli NM) is still valid. On a longer time
scale, the solution of the Markovian version of our master equation (Pauli
M) joins
the one of the non-Markovian version of our equation (Pauli NM).
\label{bisd=0.5l=0.1}}
\end{figure}

\begin{figure}[P]
\centering
\rotatebox{0}{\scalebox{0.6}{\includegraphics{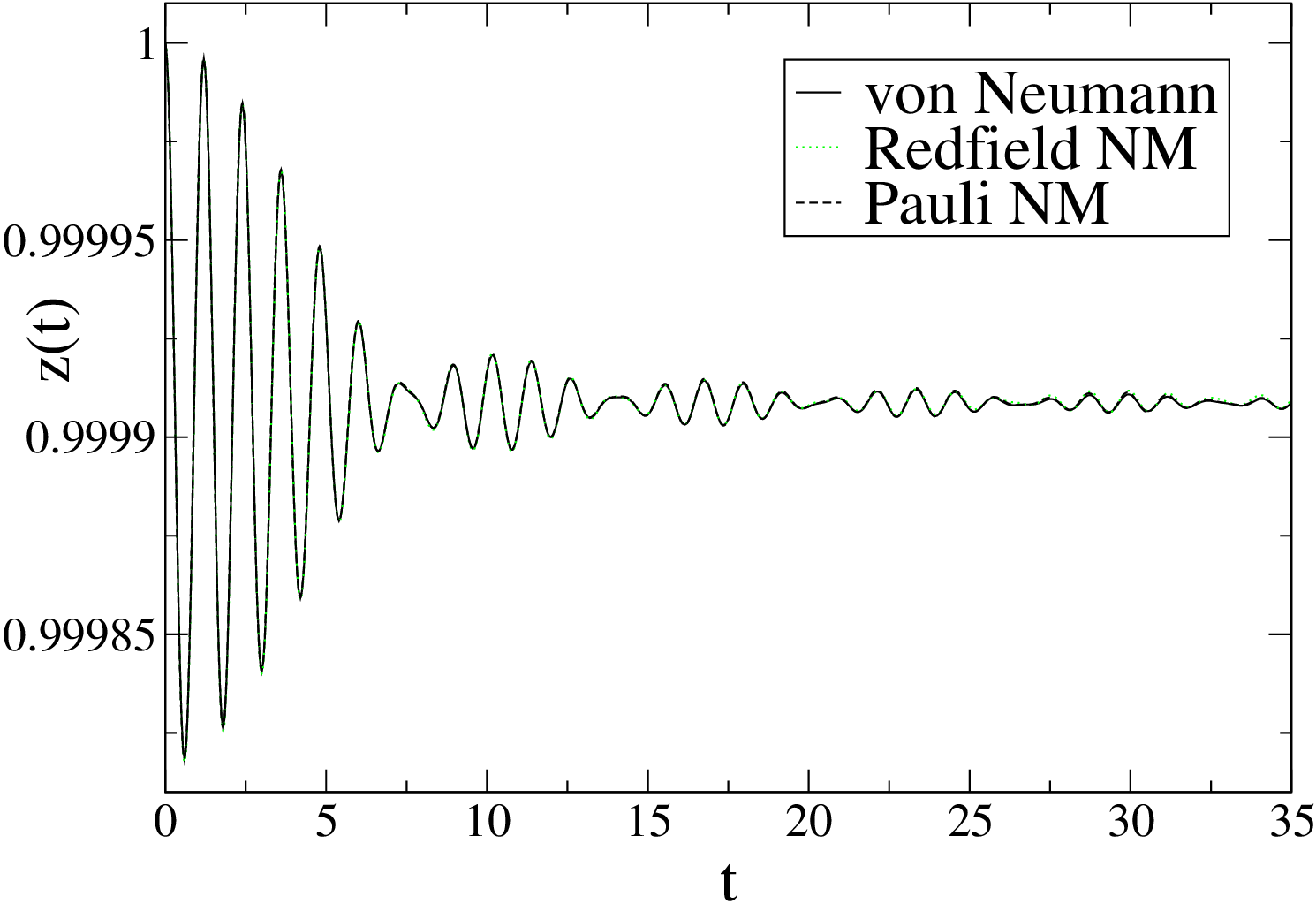}}} \\
\caption{Relaxation of the $z$ component of the spin for the
spin-GORM model for a large spin energy splitting $\Delta=5$ with
$\lambda=0.1$,
$\epsilon=0.25$, $N=2000$, and $\chi=10$. The
exact solution of the von Neumann equation is compared with the solutions
of the Pauli and Redfield non-Markovian (NM) equations.
We are in a highly non-Markovian regime. The Markovian equations are not
plotted here because their solutions are a constant equal to unity
at all times and therefore miss the whole dynamics. We see that in
this highly non-Markovian regime and on the short time scale $\tau_{corr}$
there is almost no difference
between the non-Markovian Redfield equation (Redfield NM) and our
master equation (Pauli NM). \label{z-d=5}}
\end{figure}

The results of the numerical calculation of the
time evolution of the $z$ component of the spin are depicted in
Figs. \ref{z-d=0.01}-\ref{z-d=5} for different regimes of the
spin-GORM model, i.e., for different values of the
energy splitting $\Delta$ of the two-level system
as well as of the environment energy $\epsilon$.  In all the
cases, the coupling parameter is equal to $\lambda=0.1$.

Figures \ref{z-d=0.01}, \ref{z-d=0.1} and \ref{d=0.5l=0.1}
depict the global relaxation of the $z$ component of the spin
for increasing values of the energy splitting $\Delta$.
In accordance with what we argued before on theoretical grounds,
we see in these figures that the larger the
energy splitting $\Delta$ of the system is, the
bigger is the difference between the Redfield and our master
equation. We also see
that our equation always fits very well with the exact von Neumann equation,
which is not the case of the Redfield equation. As argued before, this is due
to the fact that the Redfield equation does not take into account the
changes in the energy distribution of the environment induced by
the system transitions. When the system energy is very small this
makes almost no difference, but when it increases, this has to be
taken into account and our master equation becomes necessary.

In particular, a large discrepancy happens for the solution of
the Redfield equation in Fig. \ref{d=0.5l=0.1} although the
solution of our master equation continues to be in agreement
with the exact time evolution.  This can be understand
with Eqs. (\ref{paulispingoeMZinfini}) and
(\ref{redspingoeMZinfini}) for the asymptotic
equilibrium values of the $z$ component, which predict,
respectively,
\begin{eqnarray}
z_{\rm Pauli} &=& 0 \; , \\
z_{\rm Redfield} &=& -1 \; ,
\end{eqnarray}
for $\Delta=0.5$ and $\epsilon=-0.25$.
The discrepancy of the Redfield equation finds its origin in the
violation of energy conservation between the system
and its environment by this equation.
The Pauli equation has the advantage of allowing a correct
energy exchange between the spin and its environment,
which is crucial for obtaining the correct asymptotic equilibrium
value of the $z$ component.

We can also notice in Fig. \ref{z-d=0.01} and Figs. 
\ref{z-d=0.1}-\ref{d=0.5l=0.1} that, on the global
time scale, the non-Markovian and Markovian equations are
very close to each other. But if we look on a shorter time
scale, we see in Figs. \ref{bisz-d=0.01} and
\ref{bisd=0.5l=0.1} small differences between
the non-Markovian and Markovian equations in the early stage of the decay.
The solutions of the non-Markovian equations are in best agreement
with the exact time evolution and present a non exponential
early decay on the time scale of the environment correlation time
($\tau_{\rm corr}\simeq 10$).  In contrast, the solution of the Markovian
equation immediately enters in an exponential decay and, thus slightly
deviates from the exact solution.  This observation concerns both the
Redfield and Pauli equations.  This suggests that,
as explained in Refs. \cite{GaspRed,Suarez}, a slippage of initial conditions
is required for both Markovian equations in order to avoid
this small early-decay discrepancy.  We also observe that this discrepancy
decreases with the energy splitting $\Delta$.  This is expected
since the Markovian approximation
is valid if the time scale of the system dynamics $\frac{2
\pi}{\Delta}$ is longer than the environment time scale
($\tau_{\rm corr}\simeq 10$). Therefore, the smaller $\Delta$ is, the
better is the Markovian approximation. If one wants to make a
correct description of the system dynamics on a time scale of order
$\tau_{\rm corr}$, the non-Markovian equations should be used
(or the Markovian equations should be supplemented by a slippage of initial
conditions \cite{GaspRed,Suarez}).
We also see, in Fig. \ref{bisd=0.5l=0.1}, that the Pauli non-Markovian
equation gives
better results than the Redfield non-Markovian equation  not only on long 
time scales but also on short ones, even if
their solutions essentially coincide on a very
short time of order $\tau_{\rm corr}$ but not more.
Again, when the energy of the system is to small to affect the density of 
state of the environment, this difference between our non-Markovian equation
and the Redfield non-Markovian equation disapears (see \ref{bisz-d=0.01}
and \ref{bisd=0.5l=0.1}).

We see in Fig. \ref{z-d=5} the relaxation of the $z$ component
of the spin in a highly non-Markovian regime. The energy
difference between the two levels is here much larger than the
width of the level density of the environment. The Markovian
equations are not plotted here because they describe a constant
value equal to unity for all times and, therefore, completely miss the
dynamics. The whole spin dynamics happens on a time scale of order
$\tau_{\rm corr}$. We also see that, in this highly non-Markovian
regime, there is almost no difference between the non-Markovian
Redfield equation and our master equation on the short time scale
that we plotted. The non-Markovian Redfield equation as well as 
our non-Markovian master equation 
continue to fit with the exact dynamics even on longer time
scales.  The special structure seen in Fig. \ref{z-d=5}
can be understand by using Eqs. (\ref{redspingoeNM11})
and (\ref{redspingoeNM-1-1}).  Indeed, the curve
is the result of some time integrations
of the Bessel function $J_1(t/2)$ divided by $t$ and modulated by
$\cos\Delta t$.
Since the modulations of $\cos\Delta t$
have a period $\frac{2\pi}{\Delta}$ shorter than the
decay time $\tau_{\rm corr}\simeq 10$ of the Bessel function $J_1(t/2)$,
a shape reminiscent of a Bessel function only appears
as an envelope of the oscillations of the decay curve.

We conclude that, as expected from theoretical arguments,
our master equation gives excellent predictions, especially in situations
where the system energy is greater than or of the same order of
magnitude as the typical energy scale of variation of the
density of states of the environment. We also notice that for
non-Markovian dynamics that happen on a time scale of order
$\tau_{\rm corr}$, the non-Markovian Redfield equation give the same
result as our master equation for short time scales. But for longer
time scales our equation is the only one that correctly describes
the dynamics.

%%%%%%%%%%%%%%%%%%%%%%%%%%%%%%%%%%%%%%%%%%%%%%%%%%%%%%%%%%%%%%%%%%%%%%%%%%%%%%%%
%%%%%%%%%%%%%%%%%%%%%%%%%%%%%%%%%%%%%%%%%%%%%%%%%%%%%%%%%%%%%%%%%%%%%%%%%%%%%%%%
\section{Conclusions \label{conc}}

We derived in this paper a master equation to study the
dynamics of a quantum system interacting with its environment. This
equation is obtained by a perturbative expansion with respect to the
coupling parameter between the system and its environment.
Our equation is more general than the previously obtained  perturbative 
master equations because our equation explicitly takes into account the 
exchange of energy between the system and its environment. This effect 
is important when
the density of state of the environment varies in a significant
way on energy scales of the order of the system energy scales.

We showed how the well-known master equations of the literature
can be derived from our equation (\ref{our.nonMarkov.master.eq}) 
by performing different types of
approximations. Our equation reduces to the non-Markovian
Redfield equation (\ref{Redfield.nonMarkov.master.eq}) if one neglect 
the changes in the
density of states of the environment induced by the system.
Moreover, we showed that by performing the Markovian approximation on our
equation and neglecting the coherence contribution to the population 
dynamics, we get a Pauli-type equation (\ref{origpaulipop}) for the total
system (system $+$ environment) that describes the time evolution
in terms of distributions defined on the energy spectrum of
the environment. When one neglects the changes in the energy
distribution of the environment, the Markovian version of our
equation (that is now equivalent to the Redfield equation) reduces
to the master equation derived by Cohen-Tannoudji and corworkers
in Ref. \cite{Tannoudji}.

We have applied our equation to a two-level system interacting with a
general environment (spin-environment model), especially, in the
case where the environment operators are random matrices
(spin-GORM model). In this case, we have carried out numerical simulations of
the spin-GORM model (for which one can computed the exact
solutions) that show the greater accuracy of our master equation
with respect to the other well-known master equations in Markovian
and non-Markovian situations.

%%%%%%%%%%%%%%%%%%%%%%%%%%%%%%%%%%%%%%%%%%%%%%%%%%%%%%%%%%%%%%%%%%%%%%%%%%%%%%%%
\vspace{1cm}
\noindent{\bf Acknowledgements}

The authors thank Professor G. Nicolis for support and
encouragement in this research, as well as D. Cohen for several
fruitful discussions on the spin-GORM model. M. E. is supported by
the F.~R.~I.~A. Belgium (Fonds pour la formation \`{a} la Recherche dans
l'Industrie et
dans l'Agriculture), and P. G. by the F.~N.~R.~S. Belgium (Fonds National
pour la Recherche Scientifique).

%%%%%%%%%%%%%%%%%%%%%%%%%%%%%%%%%%%%%%%%%%%%%%%%%%%%%%%%%%%%%%%%%%%%%%%%%%%%%%%%%
\begin
{thebibliography} {1}

\bibitem{Pauli}
W. Pauli, {\it Festschrift zum 60. Geburtstage A. Sommerfelds} (Hirzel,
Leipzig, 1928).

\bibitem{Zwanzig}
R. W. Zwanzig, {\it Lectures in Theoretical Physics},
(Interscience Publishers, New York, 1961), p. 106.

\bibitem{Zubarev}
D. Zubarev, V. Morozov, and G. R\"opke, {\it Statistical Mechanics of
Nonequilibrium Processes} (Akademie Verlag, Berlin, 1996).

\bibitem{Red}
A. G. Redfield, IBM J. Res. Dev. {\bf 1}, 19 (1957).

\bibitem{Tannoudji}
C. Cohen-Tannoudji, J. Dupont-Roc, and G. Grynberg, {\it Processus
d'Interaction entre Photons et Atomes}
(CNRS Editions, Paris, 1996).

\bibitem{Gardiner}
C. W. Gardiner, P. Zoller, {\it Quantum Noise}, 2nd ed. (Spinger, Berlin, 2000).

\bibitem{Kampen}
N. G. van Kampen, {\it Stochastic Processes in Physics and Chemistry}, 2nd ed.
(North-Holland, Amsterdam, 1997).

\bibitem{Haake1}
F. Haake, {\it Statistical Treatment of Open Systems by Generalized Master
Equation}, Springer Tracts in Modern Physics, Vol. 66 (Springer, Berlin, 1973).

\bibitem{Weiss}
U. Weiss, {\it Quantum Dissipative Systems}, 2nd ed. (World Scientific, Singapore, 2000).

\bibitem{Lindblad}
G. Lindblad, Commun. Math. Phys. {\bf 48}, 119 (1976).

\bibitem{GaspRed}
P. Gaspard and M. Nagaoka, J. Chem. Phys. {\bf 111}, 5668 (1999).

\bibitem{Suarez}
A. Su\'{a}rez, R. Silbey, and I. Oppenheim, J. Chem. Phys. {\bf 97}, 5101 (1992).
 
\bibitem{Yu-Struntz} 
T. Yu, L. Di\'osi, N. Gisin, and W. Strunz, Phys. Lett. A {\bf 265}, 331 (2000).

\bibitem{Struntz} 
W. Strunz, Chem. Phys. {\bf 268}, 237 (2001).

\bibitem{Haake2} 
S. Gnutzmann and F.Haake, Z. Phys. B {\bf 101}, 263 (1996).

\bibitem{Cohen}
D. Cohen and T. Kottos, e-print cond-mat/0302319.

\bibitem{Marcus}
R. A. Marcus, Adv. Chem. Phys. {\bf 101}, 391 (1997).

\bibitem{Rice}
W. M. Gelbart, S. A. Rice, and K. F. Freed, J. Chem. Phys. {\bf 57}, 4699
(1972).

\bibitem{Legget}
A. J. Leggett, S. Chakravarty, A. T. Dorsey, M. P. A. Fisher, A. Garg, and
W. Zwerger, Rev. Mod. Phys.
{\bf 59}, 5054 (1994).

\bibitem{EspoGasp}
M. Esposito and P. Gaspard, Phys. Rev. E {\bf 68}, 066113 (2003).

\bibitem{BGS}
O. Bohigas, M. J. Giannoni, and C. Schmit, Phys. Rev. Lett. {\bf 52}, 1 (1984).

\bibitem{Guhr}
T. Guhr, A. M\"{u}ller-Groeling, and H. A. Weidenm\"{u}ller, Phys. Rep.
{\bf 299}, 189 (1998).

\bibitem{Matal}
M. L. Mehta, {\it Random Matrices and the Statistical Theory of Energy
Levels} (Academic Press, New York, 1967).

\bibitem{Porter}
C. E. Porter, {\it Statistical Theories of Spectra Fluctuations} (Academic
Press, New York, 1965).

\bibitem{Mello}
T. A. Brody, J. Flores, J. B. French, P. A. Mello, A. Pandey, and S. S. M.
Wong, Rev. Mod. Phys. {\bf 53}, 385 (1981).

\bibitem{Pereyra1}
P. A. Mello, P. Pereyra, and N. Kumar, J. Stat. Phys. {\bf 51}, 77 (1988).

\bibitem{Pereyra2}
P. Pereyra, J. Stat. Phys. {\bf 65}, 773 (1991).

\bibitem{Lutz} 
E. Lutz and H. A. Weidenm\"uller, Physica A {\bf 267}, 354 (1998).

\bibitem{Lebowitz} 
J. L. Lebowitz and L. Pastur, e-print math-ph/0307004.

\end{thebibliography}
%%%%%%%%%%%%%%%%%%%%%%%%%%%%%%%%%%%%%%%%%%%%%%%%%%%%%%%%%%%%%%%%%%%%%%%%%%%%%%%%

%%%%%%%%%%%%%%%%%%%%%%%%%%%%%%
%%%%%%%%%
%FIGURE4%
%%%%%%%%%
%%%%%%%%%%%%%%%%%%%%%%%%%%%%%%

%%%%%%%%%%%%%%%%%%%%%%%%%%%%%%
%%%%%%%%%
%FIGURE4%
%%%%%%%%%
%%%%%%%%%%%%%%%%%%%%%%%%%%%%%%

%%%%%%%%%%%%%%%%%%%%%%%%%%%%%%
%%%%%%%%%
%FIGURE4%
%%%%%%%%%
%%%%%%%%%%%%%%%%%%%%%%%%%%%%%%

%%%%%%%%%%%%%%%%%%%%%%%%%%%%%%
%%%%%%%%%
%FIGURE4%
%%%%%%%%%
%%%%%%%%%%%%%%%%%%%%%%%%%%%%%%

%%%%%%%%%%%%%%%%%%%%%%%%%%%%%%
%%%%%%%%%
%FIGURE4%
%%%%%%%%%
%%%%%%%%%%%%%%%%%%%%%%%%%%%%%%

%%%%%%%%%%%%%%%%%%%%%%%%%%%%%%
%%%%%%%%%
%FIGURE4%
%%%%%%%%%
%%%%%%%%%%%%%%%%%%%%%%%%%%%%%%

%%%%%%%%%%%%%%%%%%%%%%%%%%%%%%
%%%%%%%%%
%FIGURE4%
%%%%%%%%%
%%%%%%%%%%%%%%%%%%%%%%%%%%%%%%

%%%%%%%%%%%%%%%%%%%%%%%%%%%%%%
%%%%%%%%%
%FIGURE4%
%%%%%%%%%
%%%%%%%%%%%%%%%%%%%%%%%%%%%%%%

%%%%%%%%%%%%%%%%%%%%%%%%%%%%%%
%%%%%%%%%
%FIGURE4%
%%%%%%%%%
%%%%%%%%%%%%%%%%%%%%%%%%%%%%%%

%%%%%%%%%%%%%%%%%%%%%%%%%%%%%%
%%%%%%%%%
%FIGURE4%
%%%%%%%%%
%%%%%%%%%%%%%%%%%%%%%%%%%%%%%%

%%%%%%%%%%%%%%%%%%%%%%%%%%%%%%%%%%%%%%%%%%%%%%%%%%%%%%%%%%%%%%%%%%%%%%%%%%%%%%%%%%%
\end{document}